\documentclass[%
 reprint,
 amsmath,amssymb,
 aps,
]{revtex4-2}

\usepackage{graphicx}
\usepackage{dcolumn}
\usepackage{bm}
\usepackage{comment}
\usepackage{physics}
\usepackage{amsmath}
\usepackage{upgreek}
\usepackage{lipsum}

\usepackage{siunitx}
\DeclareSIUnit{\Torr}{Torr}
\DeclareSIUnit{\atomicpercent}{at.\%}
\DeclareSIUnit{\percent}{\%}
\DeclareSIUnit{\sccm}{sccm}
\DeclareSIUnit{\electronvolts}{eV}
\DeclareSIUnit{\ppm}{ppm}
\DeclareSIUnit{\ppb}{ppb}
\DeclareSIUnit{\count}{c}
\DeclareSIUnit{\rpm}{rpm} 

\usepackage{bm}

\usepackage{chemformula}
\setchemformula{kroeger-vink=true}

\usepackage{setspace}

\usepackage[normalem]{ulem}

\begin{document}

\preprint{APS/123-QED}
\title{Microwave-based quantum control and coherence protection of tin-vacancy spin qubits in a strain-tuned diamond membrane heterostructure}
\author{Xinghan Guo$^{1,\ast}$}
\author{Alexander M. Stramma$^{2,\ast}$}
\author{Zixi Li$^{1}$}
\author{William G. Roth$^{2}$}
\author{Benchen Huang$^{3}$}
\author{Yu Jin$^{3}$} 
\author{Ryan A. Parker$^{2}$}
\author{Jes\'{u}s Arjona Mart\'{i}nez$^{2}$}
\author{Noah Shofer$^{2}$}
\author{Cathryn P. Michaels$^{2}$}
\author{Carola P. Purser$^{2}$}
\author{Martin H. Appel$^{2}$}
\author{Evgeny M. Alexeev$^{2,4}$}
\author{Tianle Liu$^{5}$}
\author{Andrea C. Ferrari$^{4}$}
\author{David D. Awschalom$^{1,5,6}$}
\author{Nazar Delegan$^{1,6}$}
\author{Benjamin Pingault$^{6,7}$}
\author{Giulia Galli$^{1,3,6}$}
\author{F. Joseph Heremans$^{1,6}$}
\author{Mete Atat\"ure$^{2,\dagger}$}
\author{Alexander A. High$^{1,6,\dagger}$}

\affiliation{$^{1}$Pritzker School of Molecular Engineering, University of Chicago,
Chicago, IL 60637, USA}
\affiliation{$^{2}$Cavendish Laboratory, University of Cambridge, Cambridge CB3 0HE, United Kingdom}
\affiliation{$^{3}$Department of Chemistry, University of Chicago, Chicago, IL 60637, USA}
\affiliation{$^{4}$Cambridge Graphene Centre, University of Cambridge,Cambridge CB3 0FA, United Kingdom}
\affiliation{$^{5}$Department of Physics, University of Chicago, Chicago, IL 60637, USA}
\affiliation{$^{6}$Center for Molecular Engineering and Materials Science Division, 
Argonne National Laboratory, Lemont, IL 60439, USA}
\affiliation{$^{7}$QuTech, Delft University of Technology, 2600 GA Delft, The Netherlands}





\begin{abstract}
Robust spin-photon interfaces in solids are essential components in quantum networking and sensing technologies. Ideally, these interfaces combine a long-lived spin memory, coherent optical transitions, fast and high-fidelity spin manipulation, and straightforward device integration and scaling. The tin-vacancy center (SnV) in diamond is a promising spin-photon interface with desirable optical and spin properties at \SI{1.7}{\kelvin}. However, the SnV spin lacks efficient microwave control and its spin coherence degrades with higher temperature. In this work, we introduce a new platform that overcomes these challenges -- SnV centers in uniformly strained thin diamond membranes. The controlled generation of crystal strain introduces orbital mixing that allows microwave control of the spin state with \SI{99.36\pm0.09}{\percent} gate fidelity and spin coherence protection beyond a millisecond. Moreover, the presence of crystal strain suppresses temperature dependent dephasing processes, leading to a considerable improvement of the coherence time up to \SI{223\pm10}{\micro\second} at \SI{4}{\kelvin}, a widely accessible temperature in common cryogenic systems. Critically, the coherence of optical transitions is unaffected by the elevated temperature, exhibiting nearly lifetime-limited optical linewidths. Combined with the compatibility of diamond membranes with device integration, the demonstrated platform is an ideal spin-photon interface for future quantum technologies.
\end{abstract}

\maketitle

\section*{Introduction}
Color centers in diamond are a leading platform in quantum technologies, key achievements such as the demonstration of a quantum register \cite{VanDerSar2012,taminiau2014,Stas2022}, distant entanglement generation between three nodes \cite{Pompili2021}, quantum teleportation \cite{Hermans2022}, along with myriad landmarks in quantum sensing \cite{Kucsko2013,Shi2018}. In recent years, group IV centers have gained much attention due to their excellent optical properties \cite{knall2022,bhaskar2017,martinez2022,narita2023,rugar2020,Wan2020,gorlitz2020,iwasaki2017tin}. Their $D_{3d}$ symmetry renders optical transitions insensitive to first-order charge noise \cite{desantis2021,aghaeimeibodi2021,Hepp2014}. Additionally, a favorable Debye Waller factor leads to the majority of photons being emitted into the zero-phonon line, critical for spin-photon entanglement \cite{Sipahigil2016}. However, the electronic structure of group IV centers -- a spin 1/2 system with two ground state orbital branches -- renders the electron spin susceptible to phonon-driven transitions between the two branches \cite{Jahnke2015}. This temperature-dependent spin dephasing can be mitigated by operating at millikelvin temperatures \cite{becker2018,Sukachev2017} or by engineering the local phonon density of states through nanostructuring \cite{Meesala2018,Sohn2018}. Alternatively, dephasing can be mitigated by qubit engineering such as working with group IV centers with high spin-orbit coupling and thus large orbital splitting \cite{Trusheim2020}, or by leveraging spin-strain interaction in randomly-, or controllably strained group IV centers\cite{Stas2022,Sohn2018}. With a spin-orbit coupling significantly higher than those of the silicon vacancy (SiV) and the germanium vacancy (GeV) centers, the SnV center has the highest reported spin coherence time at \SI{1.7}{\kelvin} \cite{Debroux2021}. However, efficient microwave (MW) control of group IV spins requires the magnitude of spin-strain interaction to be comparable with the spin-orbit interaction, which for SnV necessitates strain approaching \SI{0.1}{\percent}. This degree of strain is challenging to achieve in microelectrical mechanical structures (MEMS) such as diamond cantilevers, with reported values on the order of \SI{0.015}{\percent} \cite{Meesala2018}. Therefore, a controlled process to generate \SI{\approx0.1}{\percent} strain in diamond is desired to improve SnV qubit performance by both increasing the operational temperature and enabling efficient MW driving.

\begin{figure*}
\centering
\includegraphics[width=2\columnwidth]{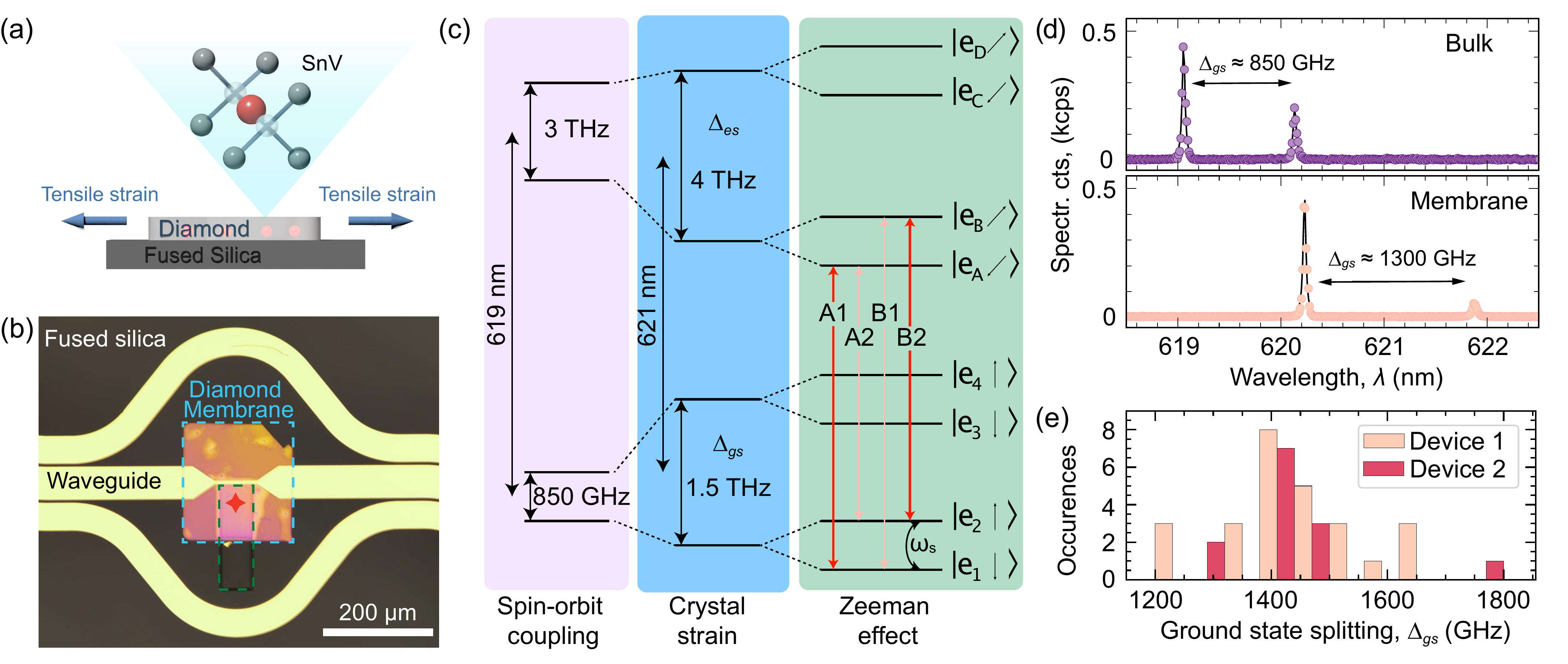}
\caption{Strained SnV in diamond membrane heterostructures. (a) Schematics of the diamond-fused silica heterostructure. The static, tensile strain inside the membrane is generated from the disparity of thermal expansion ratios of diamond and fused silica. (b) The microscope image of the diamond membrane (dashed cyan region) bonded to the fused silica substrate. A trench (dashed green region) was fabricated prior to bonding. The gold coplanar waveguide is fabricated post bonding to introduce microwave signals. The location of the SnV center used in this study is highlighted by a red star. (c) Energy level of strained SnVs. Unstrained centers, strained centers and strained centers in the presence of a magnetic field are colored in purple, blue and green, respectively. (d) The PL spectrum of a strained SnV center (orange), showing a red-shifted zero-phonon line (ZPL) wavelength with a much larger ground-state splitting compared with the values in bulk diamond (purple). (e) The statistics of the SnV ground-state splitting. Two different devices with identical layout were measured. Device 1 (orange) was used for all-optical spin control (discussed in the SI) and device 2 (purple) was used for microwave spin control.}
\label{fig1: strained SnV}
\end{figure*}

In this work, we utilize heterogeneous integration of diamond membranes to to generate strain-tuned SnVs. By bonding SnV incorporated pristine diamond membranes to a glass substrate, we leverage the heterogeneous thermal expansion coefficients of the two materials to generate a uniform, in-plane strain in the diamond to the order of \SI{0.1}{\percent}. This strain greatly increases the energy splitting between the two orbital levels of the SnV and induces orbital mixing in the spin ground state. We demonstrate MW manipulation of the spin with \SI{99.36\pm0.09}{\percent} Rabi fidelity at \SI{4.50\pm0.02}{\mega\hertz} for 24 dBm MW input power. At \SI{1.7}{\kelvin}, the implementation of dynamical decoupling allows the SnV to reach millisecond coherence time, which is largely preserved even at \SI{4}{\kelvin}, owing to the strain-induced increased ground state orbital splitting. In combination with near lifetime-limited optical linewidths up to \SI{7}{\kelvin}, our spin-photon interface is compatible with broadly utilized low-infrastructure and cost-effective portable cryogenic systems. Additionally, the demonstrated strained-membrane heterostructure maintains robustness and flexibility for additional photonic, electronic, and micro-electromechanical systems (MEMS) integration. Our SnV-based diamond membrane platform greatly reduces the technological barrier for establishing quantum nodes for networking.

\subsection*{SnVs in strained diamond}
This work relies on strain engineering to improve SnV qubit performance. First, we demonstrate that heterogeneous thermal expansion disparities between diamond and glass in a diamond-membrane heterostructure are sufficient to generate uniform strain of the magnitude necessary to beneficially impact SnV. The diamond membranes used in this work were generated via the ``smart-cut'' method combined with isotopically purified ($^{12}C$) overgrowth. The membrane thickness is nominally 150 nm, with pristine crystal quality and atomically smooth surfaces \cite{Guo2021}. To introduce a positive tensile strain inside the diamond membrane, we bond them onto \SI{500}{\micro\meter}-thick fused silica substrates---a material with a low thermal expansion coefficient ($< 1\times10^{-6}$\SI{}{\per\kelvin}) -- using a layer of hydrogen silsesquioxane (HSQ). The schematic of this strain generation method is shown in Figure \ref{fig1: strained SnV} (a). The device is then annealed at \SI{600}{\celsius}, beyond the temperature at which the HSQ solidifies to glass, bonding the heterostructure in a "zero-strain" condition \cite{Siew2000}. Due to the mismatch in thermal contraction between diamond and fused silica and the negligible thickness of the diamond membrane compared to that of the fused silica substrate, cooling down the device to cryogenic temperature regime generates a positive (tensile), static strain profile in the diamond membrane with an estimated magnitude of \SIrange{0.05}{0.1}{\percent} (see section 1.3 and 1.4 in SI for details). This passive, uniform, and membrane-compatible strain generation is complimentary to recent demonstrations of electromechanically-induced strain on suspended diamond beams \cite{Sohn2018,Dang2021}.

Figure \ref{fig1: strained SnV} (b) is the microscope image showing the layout of our diamond-membrane heterostructure device. Prior to the membrane bonding, we patterned and etched a \SI{5}{\micro\meter} deep trench on the fused silica to suspend part of the membrane and mitigate background fluorescence from the HSQ resist. To study MW control of the SnV centers, we patterned and deposited gold coplanar waveguides following membrane bonding.
 
The strain monotonically increases the orbital splitting of the SnV centers in the membranes, which can be directly verified in the photoluminescence (PL) spectra at \SI{1.7}{\kelvin}. The energy level diagram of the strained SnV is shown in Figure \ref{fig1: strained SnV} (c), highlighting the ground state orbital splitting ($\Delta_{gs}$) and the respective contributions of spin-orbit coupling, strain, and magnetic Zeeman interaction in purple, blue, and green boxes. Figure \ref{fig1: strained SnV} (d) compares the spectra of a strained (unstrained) SnV center in a diamond membrane (bulk diamond) with $\Delta_{gs} =$ \SI{\approx1300 (850)}{\giga\hertz}. This particular strained center is used in further optical, microwave and spin characterizations in this work. Remarkably, we note that all color centers in the membrane are comparably strained. As shown in Figure \ref{fig1: strained SnV} (e), we observed a distribution of the orbital branches splitting centered around \SI{1500}{\giga\hertz} across different devices with a minimum (maximum) value of \SI{1200 (1800)}{\giga\hertz}.  We carried out density functional theory (DFT) calculations to compute strain-susceptibilities and characterize  the SnV spin-strain interaction (see SI); our results show that the increase of the splitting between orbital branches from \SI{850}{\giga\hertz} to \SI{\approx1500}{\giga\hertz} due to strain, corresponds to a diamond membrane strain magnitude of \SI{0.075}{\percent}(see section 1.2 in the SI for details). The consistent strain generation, in combination with our ability to perform additional integration and nanofabrication following membrane bonding \cite{Butcher2020,Guo2023}, highlights the robustness and versatility of our platform.

\subsection*{Optical properties of SnV under strain}
\begin{figure}
\centering
\includegraphics[width=1\columnwidth]{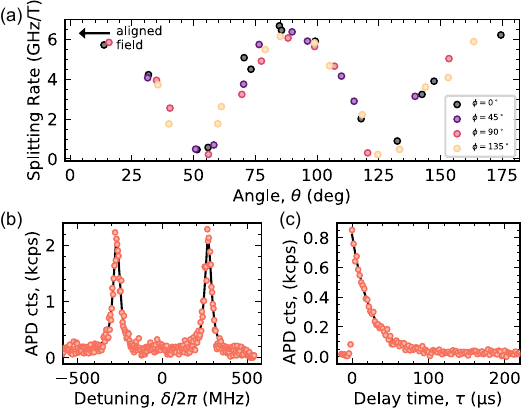}
\caption{Optical properties of the strained SnV center under applied magnetic fields at \SI{1.7}{\kelvin}. (a) The energy splitting rate between the A1-B2 spin conserving transitions with respect to the polar angle $\theta$ of the applied magnetic field at different azimuthal angle $\phi$. The aligned field is highlighted with a black arrow. (b) PLE scan, averaged over \SI{20}{\second}, of the \{A1, B2\} transitions at an aligned $B$-field with a magnitude of \SI{81.5}{\milli\tesla}. The average linewidth for both transitions are below \SI{48}{\mega\hertz}, which is less than 1.5 times of the lifetime limited value (\SI{32.26\pm0.19}{\mega\hertz}). (c) The initialization curve of the A1 transition, showing a time constant of \SI{24.2\pm0.3}{\micro\second} and an initialization fidelity of \SI{98.82}{\percent}.}
\label{fig2: optical property}
\end{figure}

To investigate the potential of strained SnV as a spin-photon interface, we first verify that the symmetry of the defect is preserved even under considerable strain by characterizing the optical transitions as a function of the magnetic ($B$) field orientation. Using the $\langle 111\rangle$ crystallographic axis -- the high symmetry axis of the SnV as the reference, we rotate the $B$ field in both polar ($\theta$) and azimuthal ($\phi$) angles at the same magnitude (\SI{0.2}{\tesla}). The absolute energy splitting between the two spin-conserving transitions (A1-B2) with respect to $\theta$ and $\phi$ is shown in Figure \ref{fig2: optical property} (a), indicating that large splittings at moderate values of magnetic field are achievable which is ideal for later SnV spin initialization and control. Similarly to the unstrained case, we observe a $\phi$ rotational symmetry of the splitting with respect to $\langle 111\rangle$, which corresponds to the intrinsic spin quantization axis. We further verify that the polarization of the SnV transitions (i.e. dipole operator matrix elements) remain along the $\langle 111\rangle$ direction (see section 3.1 of the SI), as in the unstrained case \cite{Hepp2014}.

From the $B$-field scan of the strained SnV, we note that besides the normal A1-B2 splitting maximum along the quantization axis, an additional local maximum at $\theta=$\SI{90}{\degree} -- the equator plane perpendicular to the quantization axis -- is observed, with the relative A1-B2 position being inverted, as verified by coherent population trapping measurements (see SI). This differs from the unstrained case. The novel feature arises from the moderate crystal strain (comparable in magnitude to the spin-orbit coupling) which increases the difference in effective Zeeman shift between ground and excited states, mostly visible for a magnetic field orthogonal to the spin-orbit-dictated quantization axis. As is the case for moderately strained SiV centers \cite{Sukachev2017} for MW-based control, we roughly align the $B$-field towards the quantization axis to achieve highly cycling optical transitions with cyclicity reaching $\eta\approx2500$ (see section 4.2 of SI). We note that $\eta$ can be as low as 6 when the $B$ field is perpendicular to the quantization axis, which is ideal for Raman-based all-optical control of strained SnV (see section 4.3 of SI). Moreover, by comparing the dependence on $\theta$ of the A1-B2 splitting with calculated results, we are able to determine the Stevens reduction factor $g_L$ for ground and excited states mentioned in \cite{thiering2018ab}. This model is then used to explain the optically detected magnetic resonance (ODMR) frequency of the strained SnV discussed below.

Additionally, our measurements reveal near-transform limited optical linewidths, thereby showing that the application of strain does not alter the excellent coherence properties of the optical transitions, as previously demonstrated with unstrained centers \cite{Trusheim2020,narita2023}. As shown in Figure \ref{fig2: optical property} (b), the \SI{20}{\second} average scan returns a mean linewidth of \SI{47.4\pm1.6}{\mega\hertz}, only \SI{40}{\percent} more than the lifetime-limited value of \SI{32.26\pm0.19}{\mega\hertz} (\SI{4.933\pm0.19}{\nano\second} optical lifetime, see section 3.2 of SI). The long term frequency stability of the \{A1, B2\} transitions returns a center frequency standard deviation of $\sigma_c=$\SI{23.8 \pm 0.1}{\mega\hertz} and a A1-B2 splitting standard deviation of $\sigma_s=$\SI{13.28 \pm 0.06}{\mega\hertz} (see section 3.4 of SI). This linewidth and peak stability is comparable to that of other measurements of group IV color centers in nanostructures \cite{Wan2020,Stas2022,rugar2021quantum} and thus confirms the excellent potential of these defects for quantum photonic applications.

The resolvable splitting and narrow optical transitions are crucial for the spin initialization and readout of the SnV qubit. The spin initialization curve with subtracted background is shown in Figure \ref{fig2: optical property} (c), indicating a fitted exponential decay constant of \SI{24.2\pm0.3}{\micro\second}. The initialization pulse duration was set to \SI{200}{\micro\second} allowing us to reach a fidelity of \SI{98.8}{\percent}. We note that with a cyclicity of over \SI{2500}{}, this platform is a prime candidate for single shot readout if the signal counts can be improved via on-chip structures (nanophotonics, fiber couplers or grating couplers, solid immersion lenses) \cite{Parker2023,Bhaskar2020,rugar2021quantum,fuchs2021,guney2021optimized,kuruma2021coupling} or external methods (microcavities) \cite{tomm2021,riedel2017,ruf2021}.

\subsection*{Efficient MW control of the SnV spin}
\begin{figure}[t!]
\centering
\includegraphics[width=1\columnwidth]{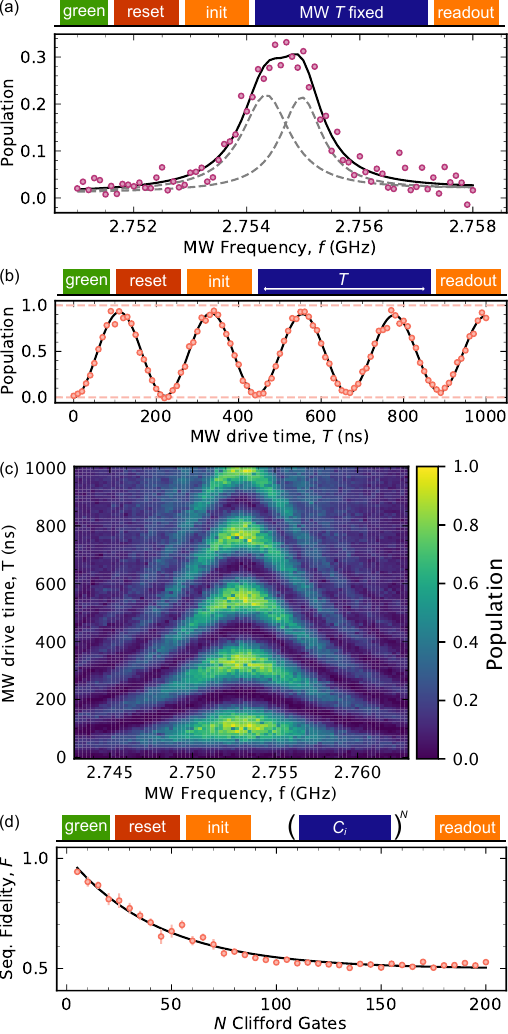}
\caption{MW control of the strained SnV center at \SI{1.7}{\kelvin}. (a) Pulsed ODMR spectrum with scanned MW frequency. The data (purple dots) is fitted with two Lorentzian functions (dashed line) split by \SI{628\pm182}{\kilo\hertz} and with a linewidth of \SI{1047\pm208}{\kilo\hertz} and \SI{891\pm197}{\kilo\hertz}, respectively. (b) Rabi oscillation of the SnV at zero detuning, indicating a Rabi frequency $\Omega/2\pi$ of \SI{4.50\pm0.02}{\mega\hertz} with a fidelity of \SI{99.36\pm0.09}{\percent}. (c) Rabi oscillation as a function of the MW driving frequency. (d) Randomized benchmarking at  \SI{1.7}{\kelvin}, showing an average gate fidelity of \SI{97.7\pm0.1}{\percent}. The Rabi frequency is set to \SI{2.8}{\mega\hertz} to avoid excess heating effects.}
\label{fig3: spin control}
\end{figure}

A critical component of a spin-photon interface is high-fidelity spin control, commonly achieved through MW driving of the electron spin. In the case of group IV centers, a MW field can only drive the spin transition in the presence of strain \cite{Pingault2017,Meesala2018}. This arises due to the orthogonality of orbital states associated with the electron spin qubit of group IV centers \cite{Hepp2014}. Strain that is comparable in strength to spin-orbit coupling relaxes this orthogonality, enabling microwave control. SnV, with larger spin-orbit coupling (\SI{850}{\giga\hertz}) and smaller strain susceptibility than SiV and GeV, requires large crystal strain to meet this criteria. This strain requirement goes beyond the achievable magnitude demonstrated via active strain tuning \cite{Meesala2018} or implantation-induced strain \cite{Stas2022}.

To demonstrate efficient MW control, we utilize the nominal \SI{0.1}{\percent} crystal strain in the diamond membrane. We estimate an effective Landé factor $g$ of \SI{1.62}{} for the transverse microwave field with the external magnetic field roughly aligned to the SnV quantization axis (see section 2.1 in SI). This value is relatively high compared with spin-orbit-dominated regime for unstrained centers (\SI{\leq0.3}{}) and is close to the free electron value ($g=2$). In addition, we tapered the MW waveguide around the measurement area by shrinking its width to \SI{6}{\micro\meter} to enhance the microwave amplitude, as shown in Figure \ref{fig1: strained SnV} (b). The distance between the target SnV and the waveguide is \SI{\approx4}{\micro\meter}, ensuring an efficient exposure to the MW driving field (see section 2.1 - 2.3 in SI for details).

We begin the MW control characterization by initializing the spin via optical pumping and scan the frequency of a MW field across the expected spin resonance while monitoring the fluorescence intensity of the spin readout at \SI{1.7}{\kelvin}. In Figure \ref{fig3: spin control} (a) we observe clear signature of optically detected magnetic resonance (ODMR) for the target SnV center. The \SI{81.5}{\milli\tesla} external magnetic field is aligned to the quantization axis by polarisation measurements and 3D field scan. The ODMR shows a profile with two overlapping peaks separated by \SI{628\pm182}{\kilo\hertz}, indicating an interaction between the electronic spin of the SnV with another system in the vicinity, likely a [\ch{^{13}C}] nuclear spin or the electron spin of a P1 center. Further investigation is needed to understand the nature of this interaction. By driving both power-broadened ODMR transitions, we are able to resonantly manipulate the spin state of the SnV with a Rabi frequency $\Omega/2\pi$ of \SI{4.50\pm0.02}{\mega\hertz}. The Rabi oscillation curve and the chevrons (Rabi oscillations with varied driving frequency) are shown in Figure \ref{fig3: spin control} (b) and (c). We observe a long-time averaged Rabi $\pi$-gate fidelity of \SI{99.36\pm0.09}{\percent}, improving significantly from previously demonstrated optical Raman-based spin control value \cite{Debroux2021}. We note that the MW power delivered to the device is approximately 24 dBm (\SI{250}{\milli\watt}) which is comparable to previous demonstrations on strained SiV \cite{Stas2022}. We also characterized the power dependence of the Rabi rate. Starting from a linear dependence, the Rabi rate deviates to sub-linear when the power surpasses 24 dBm due to excessive heating (see section 2.4 in SI), which could be optimized by replacing gold with superconducting metals (such as niobium or NbTiN) to deliver the MW signal.

We further characterize the single qubit gate fidelity of MW control via randomized benchmarking. For this, we use the following set of Clifford gates: \{$I$, $\pi_x$, $\pi_y$, $\pi_x/2$, $-\pi_x/2$, $\pi_y/2$, $-\pi_y/2$\} (see section 5.1 in SI). To prevent excessive heating effect during benchmarking which would lead to undesired spin decoherence, we apply a slightly slower Rabi rate (\SI{2.8}{\mega\hertz}, 18 dBm) which requires no time buffer between gates. The benchmarking result is shown in Figure \ref{fig3: spin control} (d). We extract an average Clifford gate fidelity of \SI{97.7\pm0.1}{\percent}, indicating power efficient MW control with high fidelity under stringent randomized benchmarking.

\subsection*{SnV spin coherence properties}
\begin{figure}
\centering
\includegraphics[width=1\columnwidth]{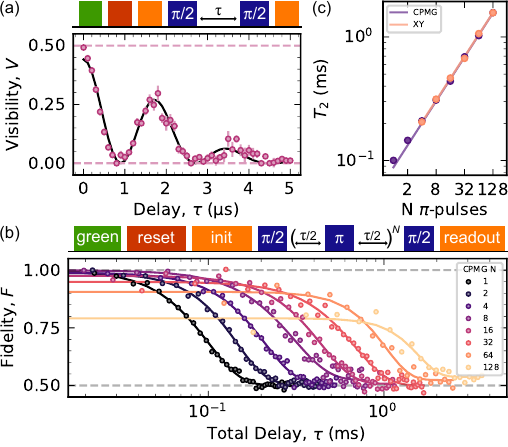}
\caption{Spin coherence of the strained SnV at \SI{1.7}{\kelvin}. (a) $T_2^*$ Ramsey of the SnV center, showing a dephasing time of \SI{2.5\pm0.1}{\micro\second}. The extra beating pattern of \SI{554\pm5}{\kilo\hertz} is estimated to be an interaction with the electron or nuclear spin in the vicinity. (b) Dynamical decoupling of the SnV via CPMG pulses. The CPMG-1 (spin-echo) returns a $T_{2,echo}$ of \SI{100\pm1}{\micro\second}, while the CPMG-128 reaches a $T_{2,CPMG128}$ of \SI{1.57\pm0.08}{\milli\second}. (c) The scaling of $T_2$ with the number of CPMG and XY pulses, showing a sub-linear dependence. }
\label{fig4: coherence}
\end{figure}

We next utilize microwave control to characterize the SnV coherence at \SI{1.7}{\kelvin}. We perform a Ramsey measurement as shown in Figure \ref{fig4: coherence} (a). The Gaussian envelope of the Ramsey oscillations corresponds to a spin dephasing time $T_2^*$ of \SI{2.5\pm0.1}{\micro\second}. Similar to ODMR, we observe interaction with a proximal spin in the Ramsey measurement, and we verify that this does not originate from the detuning of the MW signal via phase dependent readout (see section 5.2 in SI). Possible decoherence sources could be nearby vacancies and defects in the diamond membrane, as well as surface spins from both sides of the membrane \cite{sangtawesin2019}. 

Advanced pulse sequences, such as dynamical decoupling via CPMG (Carr-Purcell-Meiboom-Gill) and XY pulse sequences \cite{delange2010,souza2011}, allow us to extend the spin coherence to millisecond timescales. The CPMG results are shown in Figure \ref{fig4: coherence} (b). The $T_{2,echo}$ returns a value of \SI{100\pm1}{\micro\second}, which is already longer than \SI{35.5\pm3.0}{\micro\second} measured using all-optical spin echo process (see section 4.3 and 4.4 in SI), in the absence of optically induced dephasing mechanisms. The $T_{2,CPMG128}$, comprising 128 refocusing microwave pulses, prolongs the SnV spin coherence to \SI{1.57\pm0.08}{\milli\second}. We note that with no signal normalization being applied, the CPMG figure indicates a high signal fidelity of \SI{\approx80}{\percent} for up to 128 pulses. Future developments on the MW driving fidelity including superconducting metals and faster Rabi pulses can further improve the signal fidelity to higher numbers of pulses. We plot the relationship between the $T_2$ and the number of CPMG or XY pulses $N$ in Figure \ref{fig4: coherence} (c) and fit it with $T_2 \sim N^{\beta}$. The fitting curve returns a sub-linear dependence with a $\beta$ factor of \SI{0.593\pm0.008}{}. We observed minimal $T_2$ differences between CPMG and XY sequences. XY sequences are more resilient to control pulse errors compared to CPMG \cite{souza2011}, verifying that the observed coherence is not limited by our control (see section 5.4 in SI).

\subsection*{Spin-photon interface at \SI{4}{\kelvin}}
Finally, we demonstrate that our strained SnV platform shows state-of-the-art spin coherence for Group IV color centers at \SI{4}{\kelvin}. For Group IVs, the dominant decoherence source of the electronic spin is the electron-phonon interaction (phonon-mediated decay) between orbital branches \cite{Jahnke2015, Pingault2017}. The electron-phonon interaction rate depends on the temperature-dependent phonon population and the energy splitting $\Delta_{gs}$ between orbital branches. Therefore, enhanced coherence of the group IV centers can be achieved via either cooling down to millikelvin temperature \cite{Sukachev2017,becker2018}, increased energy splitting by using heavier group IV elements \cite{Trusheim2020}, engineering of the phonon density of states \cite{Kuruma2023}, or strain engineering \cite{Sohn2018}. Here we utilize both a heavy element (Sn as compared to Si and Ge) and crystal strain in diamond to improve electron spin coherence at elevated temperatures.

\begin{figure}
\centering
\includegraphics[width=1\columnwidth]{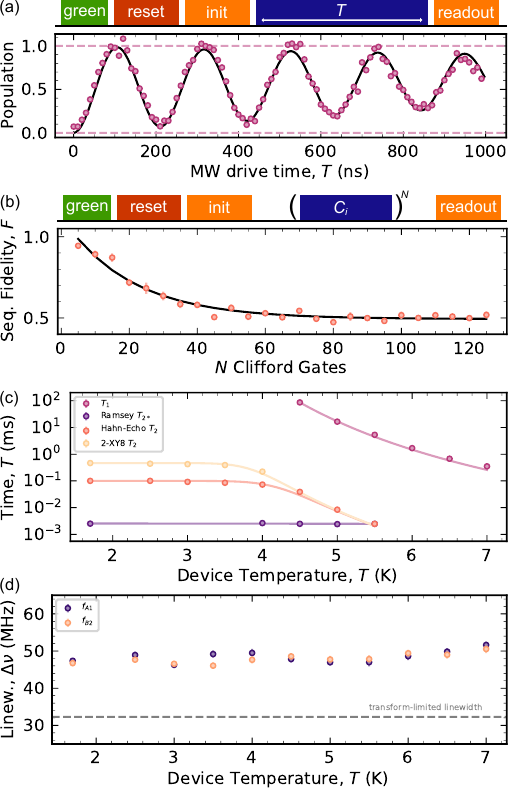}
\caption{Performance of the strained SnV center at \SI{4}{\kelvin}. (a) Rabi oscillation of the SnV center, showing a gate fidelity of \SI{97.7\pm0.5}{\percent} (b) Randomized benchmarking at \SI{4}{\kelvin}, showing an average gate fidelity of \SI{95.7\pm0.3}{\percent}. (c) Temperature dependence of the spin decay time $T_1^{\text{spin}}$, dephasing times $T_2^*$, $T_{2,\text{echo}}$, and $T_{2,\text{2XY8}}$. (d) ZPL linewidths of the two spin conserving transitions (A1, B2) with respect to the temperature, showing negligible broadening with the maximum linewidth below \SI{52\pm0.8}{\mega\hertz}. The transform-limited linewidth is shown with a dashed line.}
\label{fig5: 4K properties}
\end{figure}

The Rabi oscillation of the SnV at \SI{4}{\kelvin} is shown in Figure \ref{fig5: 4K properties} (a). The fidelity is characterized to be \SI{97.7\pm0.5}{\percent}, only slightly lower than the value at \SI{1.7}{\kelvin} due to background heating limitations. We characterize the average gate fidelity via randomized benchmarking at \SI{4}{\kelvin} using the same \SI{2.8}{\mega\hertz} Rabi rate, returning a gate fidelity of \SI{95.7\pm0.3}{\percent}, confirming the maintained high performance spin manipulation of the strained SnV at \SI{4}{\kelvin}.

Equipped with high fidelity Rabi control, we investigate the spin coherence of the SnV centers at elevated temperatures. Due to the much larger splitting $\Delta_{gs}$ of the strained SnV (\SI{\approx1300}{\giga\hertz}) compared with bulk SnV (\SI{\approx850}{\giga\hertz}), electron-phonon dephasing onsets at higher temperatures. Figure \ref{fig5: 4K properties} (c) shows the $T_1^{\text{spin}}$, $T_2^*$, $T_{2,echo}$ and $T_{2,2XY8}$ versus temperature. Fitting the same $\beta$ factor in $T_2 \sim N^{\beta}$ using Hahn-echo and XY4 coherence times returns a value of \SI{0.391\pm0.008}{} at \SI{4}{\kelvin} and \SI{0.014\pm0.000}{} at \SI{4.5}{\kelvin}, indicating that the dominant decoherence mechanism becomes phonon-induced orbital transitions instead of the spin bath. 

From Figure \ref{fig5: 4K properties} (c) we notice a much lower dephasing time compared with the decay time $T_1^{\text{spin}}$ \cite{Rogers2014}. This feature originates from the fact that only spin-flipping transitions between the lower and upper orbital branch drive $T_1^{\text{spin}}$, whereas $T_2$ is sensitive to dephasing by the spin-conserving transitions due to different precession frequencies in the orbital branches \cite{Meesala2018}. In our case, the phonon transitions are highly cycling due to the aligned magnetic field. Nevertheless, $T_2^*$ at \SI{4}{\kelvin} remains at \SI{2.7\pm0.1}{\micro\second} -- comparable to the \SI{1.7}{\kelvin} value, and $T_{2,echo}$ only decreases slightly to \SI{74\pm2}{\micro\second}, with $T_{2,2XY8}$ reaching the depolarization-limited $T_2$ -- \SI{223\pm10}{\micro\second}. It is worth emphasizing that all of these are record high values for all group IV spin qubits at \SI{4}{\kelvin} to date.

To demonstrate the potential of the strained SnV center as a promising spin-photon interface at elevated temperature, we investigate the temperature dependence of the SnV optical coherence. As shown in Figure \ref{fig5: 4K properties} (d), we observe that the ZPL linewidth remains unchanged for both A1 and B2 transitions up to \SI{7}{\kelvin} with the maximum linewidth remaining below \SI{52\pm0.8}{\mega\hertz}---only \SI{60}{\percent} higher than lifetime-limited values. In the future, modest Purcell enhancement of SnV emission rates with on-chip nanophotonics or microcavities can generate fully lifetime-limited photons suitable for efficient entanglement generation.

\section*{Conclusions}
In this work, we demonstrate that SnV in strained diamond membranes is a promising platform for quantum technologies. We create simple heterostructures that leverage differences in thermal expansion to passively generate significant strain of \SI{0.05}{\percent} to \SI{0.1}{\percent} in diamond, enabling efficient, high fidelity microwave control of the SnV spin. The presence of the strain also suppresses the phonon-mediated decay and improves the spin coherence of the SnV at \SI{4}{\kelvin}, which greatly reduces the technological barrier for quantum networking applications. We reach a Rabi $\pi$ gate fidelity of  \SI{99.36\pm0.09}{\percent}  (\SI{97.7\pm0.5}{\percent}) with a randomized single qubit gate fidelity of \SI{97.7\pm0.1}{\percent} (\SI{95.7\pm0.3}{\percent}) at \SI{1.7}{\kelvin} (\SI{4}{\kelvin}). Dynamical decoupling sequences allow the SnV spin coherence to reach \SI{1.57\pm0.08}{\milli\second} at \SI{1.7}{\kelvin} and \SI{223\pm10}{\micro\second} at \SI{4}{\kelvin}. In the future this value can be further enhanced by generating higher strain through heterostructure optimization and/or additional active tuning. Our platform, derived from scalable diamond membrane generation, is compatible with further on-chip integration, such as microwave coplanar waveguides, integrated photonics \cite{Guo2023}, and MEMS. Finally, \SI{4}{\kelvin} cryostats are relatively affordable and less infrastructure-intensive in comparison to cryogen-free \SI{1.7}{\kelvin} and \SI{}{\milli\kelvin} dilution-fridge systems. Therefore, the demonstrated spin-photon interface at \SI{4}{\kelvin} can reduce barriers to widespread utilization and deployment of solid-state quantum technologies.

\section*{Acknowledgements}

This work on strain engineering of Group IV color centers is supported by the Air Force Office of Scientific Research under award number FA9550-22-1-0518. This work acknowledges funding through Q-NEXT, supported by the U.S. Department of Energy, Office of Science, National Quantum Information Science Research Centers. The experiment receives support from the ERC Advanced Grant PEDESTAL (884745) and the EU Quantum Flagship 2D-SIPC. Membrane integration research is supported by NSF award AM-2240399. Diamond growth related efforts were supported by the U.S. Department of Energy, Office of Basic Energy Sciences, Materials Science and Engineering Division (N.D.). The membrane bonding work is supported by NSF award AM-2240399. This work made use of the Pritzker Nanofabrication Facility (Soft and Hybrid Nanotechnology Experimental Resource, NSF ECCS-2025633) and the Materials Research Science and Engineering Center (NSF DMR-2011854) at the University of Chicago. A.M.S. acknowledges support from EPSRC/NQIT, R.A.P. from the General Sir John Monash Foundation and G-research, J.A.M. from the Winton Programme and EPSRC DTP, C.P.M. from the EPSRC DTP. B. P. acknowledges funding from the European Union’s Horizon 2020 research and innovation programme under the Marie Skłodowska-Curie Grant Agreement No. 840968. The authors thank Eric I. Rosenthal, Abigail J. Stein, Hope Lee, Srujan Meesala and Dorian Gangloff for insightful discussions, Haoxiong Yan and Ming-Han Chou for experimental help.

\textit{Note added} -- During the completion of the work, we became aware of another related manuscript \cite{rosenthal2023}. 

${}^{*}$ These authors contributed equally to this work.

${}^{\dagger}$ Correspondence should be addressed to: \href{mailto:ma424@cam.ac.uk}{ma424@cam.ac.uk}, \href{mailto:ahigh@uchicago.edu}{ahigh@uchicago.edu}.

\section*{Competing interest}
A. A. H., X. G., Z. L., T. L., N. D., and F. J. H. filed a provisional patent for the strain generation of bonded membranes.

\end{document}


\title{Supplementary information: Microwave-based quantum control and coherence protection of tin-vacancy spin qubits in a strain-tuned diamond membrane heterostructure} 
\author
{Xinghan Guo$^{1,\dagger}$, Alexander M. Stramma$^{2,\dagger}$, Zixi Li$^{1}$, William G. Roth$^{2}$, \\
Benchen Huang$^{3}$, Yu Jin$^{3}$, Ryan A. Parker$^{2}$, Jesús Arjona Martínez$^{2}$,\\
Noah Shofer$^{2}$, Cathryn P. Michaels$^{2}$, Carola P. Purser$^{2}$, Martin H. Appel$^{2}$,\\
Evgeny M. Alexeev$^{2,4}$,Tianle Liu$^{5}$,
Andrea C. Ferrari$^{4}$, David D. Awschalom$^{1,5,6}$, \\ 
Nazar Delegan$^{1,6}$, Benjamin Pingault$^{6,7}$, 
Giulia Galli$^{1,3,6}$, \\
F. Joseph Heremans$^{1,6}$, Mete Atatüre$^{2,\ast}$, Alexander A. High$^{1,6,\ast}$\\
\textit{\normalsize{$^{1}$Pritzker School of Molecular Engineering, University of Chicago,
Chicago, IL 60637, USA}}\\
\textit{\normalsize{$^{2}$Cavendish Laboratory, University of Cambridge, Cambridge CB3 0HE, United Kingdom}}\\
\textit{\normalsize{$^{3}$Department of Chemistry, University of Chicago, Chicago, IL 60637, USA}}\\
\textit{\normalsize{$^{4}$Cambridge Graphene Centre, University of Cambridge, }}\\
\textit{\normalsize{Cambridge CB3 0FA, United Kingdom }}\\
\textit{\normalsize{$^{5}$Department of Physics, University of Chicago, Chicago, IL 60637, USA}}\\
\textit{\normalsize{$^{6}$Center for Molecular Engineering and Materials Science Division, }}\\
\textit{\normalsize{Argonne National  Laboratory, Lemont, IL 60439, USA}}\\
\textit{\normalsize{$^{7}$ QuTech, Delft University of Technology, 2600 GA Delft, The Netherlands}} \\
\textit{\normalsize{$^{\dagger}$ These authors contributed equally to this work.}}\\
\textit{\normalsize{$^\ast$E-mail: ma424@cam.ac.uk}}\\
\textit{\normalsize{$^\ast$E-mail: ahigh@uchicago.edu}}
}

\date{}
\maketitle 
\section{Tin vacancy center (\ch{SnV-}) in strained diamond membranes}
\subsection{Hamiltonian of the strained \ch{SnV-}}
\label{section:Hamiltonian}
The \ch{SnV-} center is a spin-$1/2$ system. In a mean-field orbital picture, the system has three electrons in four spin orbitals ($\{\ket{e_x \uparrow}, \ket{e_x \downarrow}, \ket{e_y \uparrow}, \ket{e_y \downarrow}\}$). Both its electronic ground and excited states are doubly-degenerate; the degeneracy may be lifted by applying strain and/or by spin-orbit interaction.  We write the spin Hamiltonian of the system in the minimum model of 4 electrons and 3 orbitals, for the ground ($g$) and excited ($u$) state $H_{g, u}$, as the sum of four terms: spin-orbit (SO) interaction ($\hat{H}_{\text{SO}}$); electron-phonon interaction due to the Jahn-Teller effect; strain field, and interaction with an external, static magnetic field $B$ (Zeeman effect, $\hat{H}_{Z}$). Following Ref~\cite{Hepp2014}, we write the term arising from Jahn-Teller distortions in the same form as that describing the strain interaction. Below we merge the two terms into one, that for simplicity we call $\hat{H}_{\text{strain}}$. Hence the Hamiltonian is written as:
\begin{equation}
    \hat{H}_{\text{sys}} = \hat{H}_{\text{SO}} + \hat{H}_{\text{strain}} + \hat{H}_{Z}. \label{eq:system_hamiltonian}
\end{equation}
In the following three subsections, we discuss each term of the Hamiltonian.

\subsubsection{Spin-orbit coupling}
The component of the orbital angular momentum operator $\hat{L}_x, \hat{L}_y$ vanish for the Hamiltonian expressed in the $\{\ket{e_x}, \ket{e_y}\}$ basis~\cite{Hepp2014} and only the following term is non-zero:
\begin{equation}
    \hat{L}_z = \left[\begin{matrix}
    0 & -i\\
    i & 0
    \end{matrix}\right],
\end{equation}
where we have set $\hbar$ to 1. Therefore, using the $\{\ket{e_x \uparrow}, \ket{e_x \downarrow}, \ket{e_y \uparrow}, \ket{e_y \downarrow}\}$ basis, the SO Hamiltonian can be represented as:
\begin{align}
    \hat{H}_{SO}=\lambda\hat{L}_z\hat{S}_z=\frac{\lambda}{2}
    \left[\begin{matrix}
    0 & -i\\
    i & 0
    \end{matrix}\right] \otimes
    \left[\begin{matrix}
    1 & 0\\
    0 & -1
    \end{matrix}\right]=
        \left[\begin{matrix}
    0 & 0 & -i\lambda/2 & 0\\
    0 & 0 & 0 & i\lambda/2\\
    i\lambda/2 & 0 & 0 & 0\\
    0 & -i\lambda/2 & 0 &0
    \end{matrix}\right].  
\end{align}

\subsubsection{Strain field}
The term of the Hamiltonian representing the presence of a strain field can be written as:
\begin{align}\label{strainhamiltonian}
    \hat{H}_{\text {strain}}=\left[\begin{array}{cc}
\varepsilon_{A_{1}}-\varepsilon_{E_{x}} & \varepsilon_{E_{y}} \\
\varepsilon_{E_{y}} & \varepsilon_{A_{1}}+\varepsilon_{E_{x}}
\end{array}\right] \otimes \mathbb{I}_{2}.
\end{align}
The elements $\{\varepsilon_{A_1}, \varepsilon_{E_x}, \varepsilon_{E_y}\}$ represent the energy response induced by strain belonging to the different irreducible representations $A_1, E_x, E_y$ of the $D_{3d}$ point group of the defect, and are expressed in the \ch{SnV-} center's local frame, where the $z$-axis corresponds to the high symmetry axis of the \ch{SnV} which is the quantization axis. For example,  $\varepsilon_{A_1} = \braket{\Psi|(H - H_0)|\Psi}$, where $H_0$ is the electronic Hamiltonian in the absence of strain and $H$ is the electronic Hamiltonian, which includes the strain field applied to the supercell by changing the lattice parameters. Here $\ket{\Psi}$ represents a Slater determinant expressed in the $\{\ket{e_x \uparrow}, \ket{e_x \downarrow}, \ket{e_y \uparrow}, \ket{e_y \downarrow}\}$ basis.

We can write each term of Eq. \ref{strainhamiltonian} as a linear combination of the components of the strain tensor ($\epsilon$):
\begin{equation}
\begin{split}
    \varepsilon_{A_{1}} & = t_{\perp} \left(\epsilon_{xx} + \epsilon_{yy} \right) + t_{\|} \epsilon_{zz},\\
    \varepsilon_{E_{x}} & = d\left(\epsilon_{xx}-\epsilon_{yy}\right) + f \epsilon_{zx},\\
    \varepsilon_{E_{y}} & = -2d \epsilon_{xy} + f \epsilon_{yz},
\end{split}
\end{equation}
where $\epsilon_{xx}, \epsilon_{yy}, \epsilon_{zz}$ represent the diagonal components of the strain tensor in the $x, y, z$ directions and $\epsilon_{xy}, \epsilon_{yz}, \epsilon_{zx}$ represent the shear strain components; $t_{\perp}, t_{\|}, d$, and $f$ are partial derivatives written as $\frac{\partial \varepsilon_{A_1}}{\partial (\epsilon_{xx} + \epsilon_{yy})}, \frac{\partial \varepsilon_{A_1}}{\partial \epsilon_{zz}}, \frac{\partial \varepsilon_{E_x}}{\partial (\epsilon_{xx} - \epsilon_{yy})}, \frac{\partial \varepsilon_{E_x}}{\partial \epsilon_{zx}}$, respectively. These four strain-susceptibility parameters completely describe the strain-response of the ground and excited electronic states. In the following,  we ignore the diagonal term $\epsilon_{A_{1}}$, which amounts to  a global emission wavelength shift. Hence, the strain Hamiltonian has the following form: 
\begin{align}
    \hat{H}_{\text {strain }}=\left[\begin{array}{cc}
-\varepsilon_{E_{x}} & \varepsilon_{E_{y}} \\
\varepsilon_{E_{y}} & \varepsilon_{E_{x}}
\end{array}\right] \otimes \mathbb{I}_{2} = \left[\begin{array}{cccc}
-\varepsilon_{E_{x}} & 0 & \varepsilon_{E_{y}} &0 \\
0 & -\varepsilon_{E_{x}} & 0 & \varepsilon_{E_{y}}\\
\varepsilon_{E_{y}} &0 & \varepsilon_{E_{x}} & 0\\
0 & \varepsilon_{E_{y}} & 0 & \varepsilon_{E_{x}}
\end{array}\right]. \label{eq:strain_hamiltonian}
\end{align}

\subsubsection{Zeeman effect} 
\label{subsection: Zeeman}
Due to the $D_{3d}$ symmetry of the defect, the orbital component $H_{Z,L}$ of the Hamiltonian  $H_{Z}$ only includes a term  $\hat{L}_z B_z$~\cite{Hepp2014}, with a pre-factor $q$~\cite{thiering2018ab}, called in the literature  {\it effective reduction factor}, accounting for: (i) electron-phonon interaction (so-called Ham term), and  (ii) the symmetry of the defect being lower than $O(3)$ (so-called Steven's factor). Note that both terms have different values for the ground and excited states and hence the $q$ parameter is different in the ground and excited states. 
The $H_{Z}$ Hamiltonian is written as the sum of an orbital $H_{Z,L}$ and spin component $H_{Z,S}$
\begin{equation} \label{eq:Zeeman}
    \hat{H}_Z =\hat{H}_{Z,L}+\hat{H}_{Z,S}= q \mu_B\gamma_L \hat{L}_z B_z + g\mu_B \hat{\textbf{S}} \cdot \textbf{B} - 2\mu_B\delta_f \hat{S}_z B_z,
\end{equation}
where $\mu_B$ is the Bohr magneton and $B_x, B_y, B_z$ are the components of the external, static magnetic field along the crystal frame $x, y, z$ directions, respectively. The last term on the right hand side of Eq. \ref{eq:Zeeman} originates from correcting with a factor $\delta$ the electronic Landé $g$ factor to account for spin-phonon interaction mediated by spin-orbit coupling~\cite{thiering2018ab}. For all the experimental interpretations, we'll only use the Ham factor and set the Steven's factor as $1$ except section~\ref{section:optical splitting with B field} where we gave estimates on the possible values of Steven's factor.

\subsection{Strain susceptibility}
\label{section: strain susceptibility}
In the presence of a strain field, the degeneracy of the ground ($gs$) and excited ($es$) states is lifted and we call $\triangle_{\text{gs}(\text{es})}$ the energy difference between the two states split by the degeneracy. By diagonalizing the strain Hamiltonian defined in Eq.~\ref{eq:strain_hamiltonian}, we obtain:
\begin{equation}
    \triangle_{\text{gs}(\text{es})} = 2\sqrt{[d_{\text{gs}(\text{es})}(\epsilon_{xx} - \epsilon_{yy}) + f_{\text{gs}(\text{es})}\epsilon_{zx}]^2 + [-2d_{\text{gs}(\text{es})}\epsilon_{xy} + f_{\text{gs}(\text{es})}\epsilon_{yz}]^2},
    \label{eq:strain_splitting}
\end{equation}
where the strain-susceptibilities are computed from density functional theory (DFT) calculations. We performed DFT calculations employing both the PBE~\cite{perdew1996generalized} and SCAN~\cite{sun2015strongly} functionals, and a 511-atom supercell with a [0.5, 0.5] occupation number for the $\ket{e_x \downarrow}, \ket{e_y \downarrow}$ orbitals. We approximated the splittings by the energy difference of the corresponding Kohn-Sham (KS) orbitals. The strain susceptibilities $d_{\text{gs}}, d_{\text{es}}, f_{\text{gs}}, f_{\text{es}}$ can be obtained from Eq.~\ref{eq:strain_splitting} by varying the lattice parameters of the supercell to generate $(\epsilon_{xx} - \epsilon_{yy})$ and $\epsilon_{zx}$ strain, respectively. Our results are summarized in Table.~\ref{tab:strain_susceptibility}. Note the similarity of results obtained with the two different functionals.

\begin{table}[h!]
\centering
\begin{tabular}{c c c c c}
 \hline
 Functional & $d_{\text{gs}}$ & $d_{\text{es}}$ & $f_{\text{gs}}$ & $f_{\text{es}}$ \\
 \hline
 PBE & 0.787 & 0.956 & -0.562 & -2.555 \\ 
 SCAN & 0.834 & 0.921 & -0.563 & -2.592 \\
 \hline
\end{tabular}
\caption{Computed strain susceptibilities (see text) of the SnV$^-$ defect in diamond, in units of  PHz/strain, obtained with the PBE and SCAN functionals.}
\label{tab:strain_susceptibility}
\end{table}

\subsection{Strain magnitude simulation}
We use COMSOL to simulate the strain profile of the suspended area measured in experiments. Since the strain expression $\epsilon$ in section \ref{section:Hamiltonian} is defined in local \ch{SnV} frame while the simulation result $\Tilde{\epsilon}$ returns to the lab frame, a combination of rotation matrices are applied. From $\braket{100}$ to $\braket{110}$ to $\braket{111}$, the rotation operators are $\hat{R}_z(45^\circ)$ and $\hat{R}_y(54.7^\circ)$, respectively:
\begin{align}
    \epsilon = \hat{R}^{\dagger}_y(54.7^\circ) \hspace{1mm} \hat{R}^{\dagger}_z(45^\circ) \hspace{1mm} \Tilde{\epsilon} \hspace{1mm} \hat{R}_z(45^\circ) \hspace{1mm} \hat{R}_y(54.7^\circ)
\end{align}
Here $\hat{R}_y(\theta)$ and $\hat{R}_z(\theta)$ refer to: 
\begin{align}
\label{eq:rotation}
    \hat{R}_y(\theta) = \left[\begin{matrix}
    \text{cos}(\theta) & 0 & \text{sin}(\theta)\\
    0 & 1 & 0\\
    -\text{sin}(\theta) & 0 & \text{cos}(\theta)
    \end{matrix}\right], \hspace{10mm}
    \hat{R}_z(\theta) = \left[\begin{matrix}
    \text{cos}(\theta) & -\text{sin}(\theta) & 0\\
    \text{sin}(\theta) & \text{cos}(\theta) & 0\\
    0 & 0 & 1
    \end{matrix}\right]
\end{align}

\begin{figure}[ht]
\centering
\includegraphics[width=0.9\textwidth]{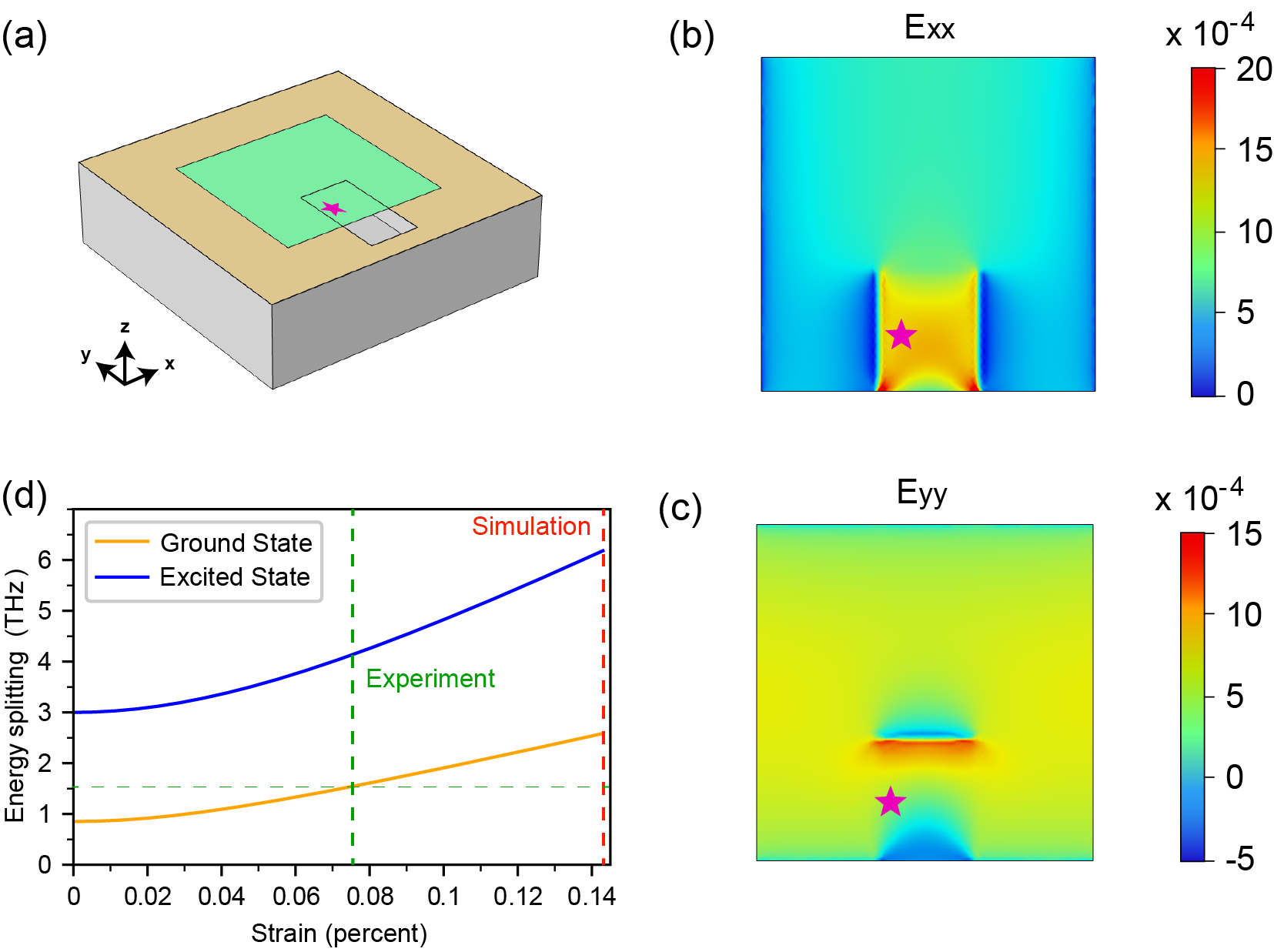}
\caption{Strain profile of the diamond membrane. (a) The 3D structure of the simulated device. The total length of the carrier wafer is limited to 350 um. (b-c) The $E_{xx}$ ($E_{yy}$) strain profile on the membrane. The position of the SnV center characterized in the main text and SI is labelled as a red star. (d) The energy splitting between two orbital branches in ground and excited states. The simulated strain and the experimentally observed strain magnitudes are labelled in dashed red and green lines, respectively. }
\label{subfig:COMSOL strain}
\end{figure}

In COMSOL simulation, we use the actual three dimensional (3D) geometry for the diamond membrane and the trench. The temperature-dependent thermal expansion ratio for diamond and fused silica are obtained from these references:\cite{hahn1972thermal, oikawa1999thermal, reeber1996thermal, mcskimin1972elastic}. The initial strain-free temperature is set to \SI{450}{\celsius} which is the HSQ healing temperature \cite{siew2000thermal}, while the final temperature is set to \SI{4}{\kelvin}. We note that thermal expansion ratios for both fused silica and diamond become negligible below \SI{30}{\kelvin}, thus the simulated strain profile is nearly constant within the temperature range of interest (\SIrange{1.7}{7}{\kelvin}) in this study. The simulated structure and the strain distributions of $E_{xx}$ and $E_{yy}$ are shown in Figure \ref{subfig:COMSOL strain}. Since the off-diagonal shear strain is 2-3 orders of magnitude smaller than the diagonal tensile strain, we use the following matrix to represent the simulated strain value:
\begin{align}
    \Tilde{\epsilon} = \left[\begin{matrix}
    1.3e^{-3} & 0 & 0\\
    0 & 6.8e^{-4} & 0\\
    0 & 0 & -2.5e^{-4}
    \end{matrix}\right], \hspace{10mm}
    \epsilon = \left[\begin{matrix}
    1.6e^{-4} & -1.8e^{-4} & 5.8e^{-4}\\
    -1.8e^{-4} & 9.9e^{-4} & -2.5e^{-4}\\
    5.8e^{-4} & -2.5e^{-4} & 5.8e^{-4}
\end{matrix}\right]
\end{align}
We note that although $\Tilde{\epsilon}$ only includes diagonal elements, the transformed strain tensor $\epsilon$ in SnV local frame contains non-negligible off-diagonal elements which could affect the properties of the SnV center through both $d$ and $f$ parameters. By comparing the simulated branch splitting value using equation \ref{eq:strain_splitting} and PBE results with the actual experimental values, we observed the actual strain to be 0.55 times the simulated value, as shown in Figure \ref{subfig:COMSOL strain} (d). This magnitude mismatch could come from either the approximation of energy splittings from KS orbitals being inaccurate, the mismatch of the thermal expansion ratios between COMSOL simulation and reality, or an even lower softening temperature of HSQ rather than the healing temperature \cite{siew2000thermal}. More comprehensive studies of the energy response to strain would require a higher level of method, e.g., embedding theory~\cite{sheng2022green}, which we left for future investigations. In the following calculations, we add this 0.55 pre-factor to the simulated strain tensor to best capture the system properties.

\subsection{Strain magnitude discussion}
\begin{figure}[ht]
\centering
\includegraphics[width=0.5\textwidth]{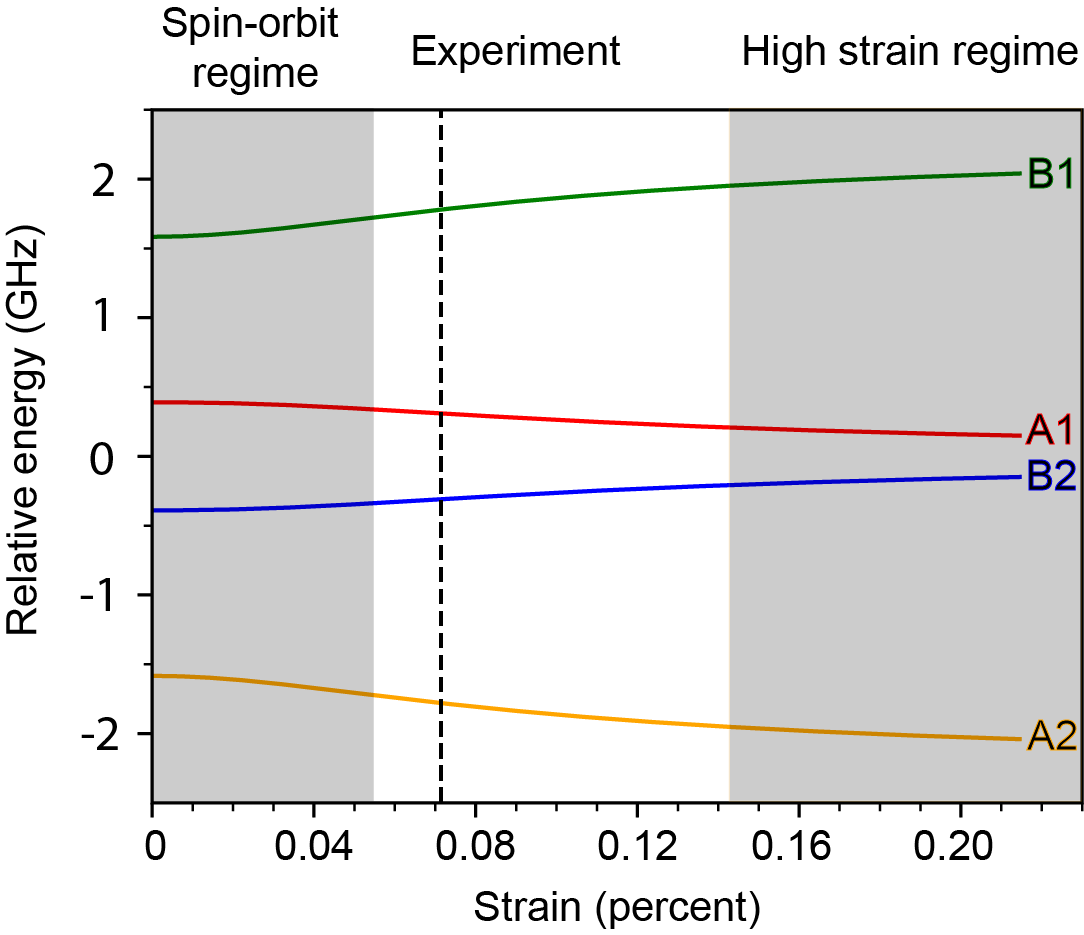}
\caption{Spin-conserving transition frequencies with respect to different strain magnitudes. The relative strain profile stays the same with only the scaling factor sweeping. The magnitude of the strain is defined by the norm of the tensor. The $B$ field is set to be \SI{80}{\milli\tesla} along the quantization axis $\braket{111}$ which is in line with the experimental configuration. }
\label{subfig:spin conserving}
\end{figure}

We qualitatively categorize the strain magnitude to different regimes via the ground state energy splitting $\Delta_{gs}$. In the spin-orbit regime, this energy splitting is nearly constant, while the splitting is linear with the external strain when in high-strain regime. Guided by that, we use $\Delta_{gs}=\SI{1200}{\giga\hertz}$ as the boundary between the spin-orbit regime and the intermediate regime, and $\Delta_{gs}=\SI{2600}{\giga\hertz}$ to identify intermediate and high strain regime. For our work, those values corresponding to strain magnitudes of \SI{0.055}{\percent} and \SI{0.143}{\percent}. Here to plot the optical transitions, the magnetic field is set to \SI{80}{\milli\tesla} along the quantization axis ($\braket{111}$ direction) and the strain profile is set to be the same as COMSOL simulated profile but with an additional scaling factor. The calculated relative energy difference of the four \{$A1, A2, B1, B2$\} transitions are plotted in Figure \ref{subfig:spin conserving}. Compared with the defined low strain ``spin-orbit'' regime and the high strain regime, our experimentally observed result sits in between, indicating a non-trivial intermediate region where neither spin-orbit coupling or strain shall we treated as perturbation terms. We note that unlike SiV centers \cite{meesala2018strain}, SnV obtains different quenching $q$ factors at ground (0.471) and excited states (0.125) \cite{thiering2018ab}, leading to a non-zero splitting between spin-conserving transitions. We also extrapolate a qubit frequency $\omega_s$ of \SI{2.1}{\giga\hertz} which is lower than the ODMR frequency reported in the main text (\SI{2.755}{\giga\hertz}). This mismatch could originate from the slight difference between the displayed and the real magnetic field due to hysteresis and the deviation of the effective reduction factors under strain from that in Ref~\cite{thiering2018ab}.

\section{MW control of the SnV}
\subsection{MW magnetic response}
At zero strain, the SnV spin qubit cannot be manipulated by microwave due to different orbits associated with the spin level. As stated in \cite{meesala2018strain}, the presence of the crystal strain introduces orbit superposition to SnV's spin qubit eigenstates, allowing for the coherent control of SnV via external microwave field with qubit frequency $\omega_s$. Here we use the electronic $g$ factor to characterize the ability of the MW field (AC B filed) to the spin state of the SnV, including both spin and orbit response of the external magnetic field:
\begin{align}
    g = \frac{2}{\mu_{B}}\bra{e1\downarrow}(\hat{H}_{Z,L}^{ac}+\hat{H}_{Z,S}^{ac})\ket{e2\uparrow}
\end{align}   
\begin{align}
    \hat{H}_{Z,L}^{ac}+\hat{H}_{Z,S}^{ac} = \left[\begin{matrix}
    B_{z}^{ac} & B_{x}^{ac}-i B_{y}^{ac} & -i q B_{z}^{ac} & 0 \\
    B_{x}^{ac}+i B_{y}^{ac} & -B_{z}^{ac} & 0 & -i q B_{z}^{ac} \\
    i q B_{z}^{ac} & 0 & B_{z}^{ac} & B_{x}^{ac}-i B_{y}^{ac} \\
    0 & i q B_{z}^{ac} & B_{x}^{ac}+i B_{y}^{ac} & -B_{z}^{ac}
    \end{matrix}\right]
\end{align}

\begin{figure}[ht]
\centering
\includegraphics[width=0.7\textwidth]{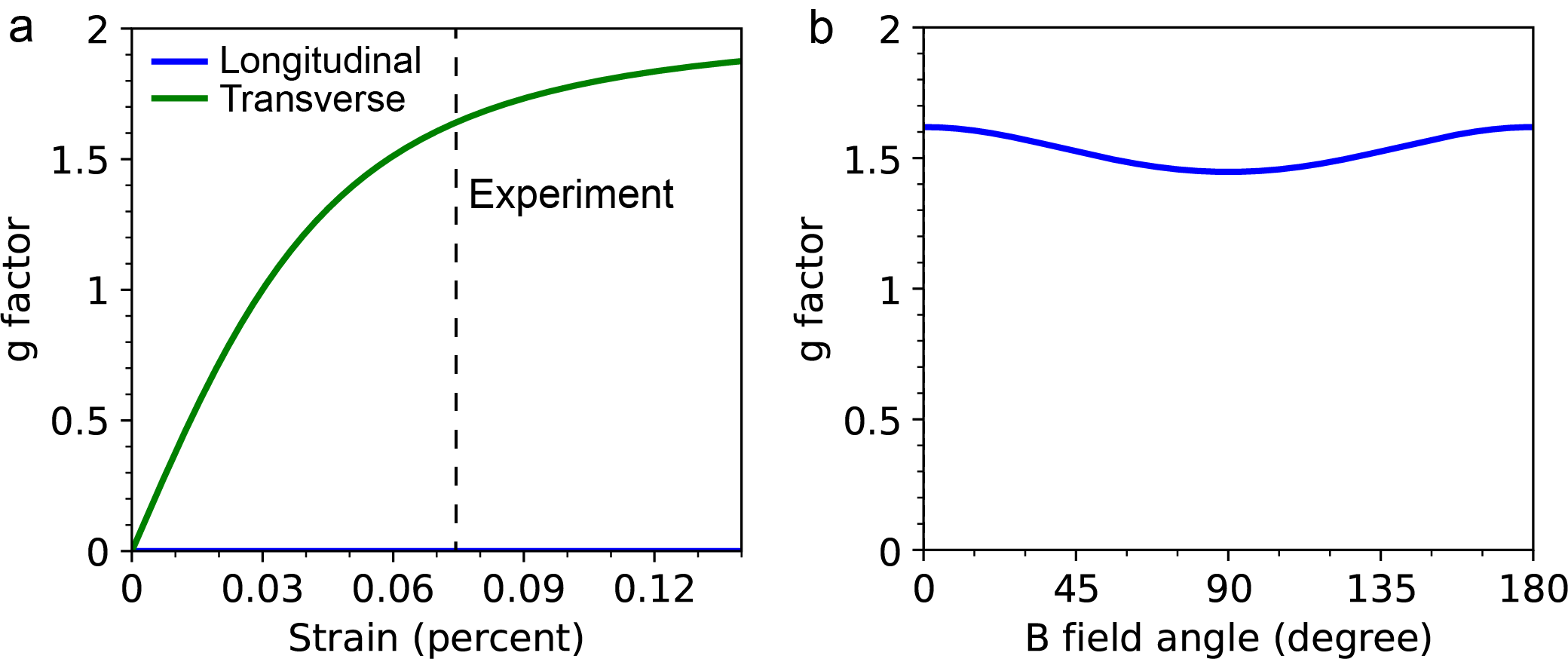}
\caption{MW $g$ factor calculation. (a) The transverse and longitudinal $g$ factor with respect to the strain magnitude. (b) The angular dependence of the $g$ factor. The static B field is set to be \SI{80}{\milli\tesla} along the quantization axis $\braket{111}$. }
\label{subfig:MW g factor}
\end{figure}

Here the $\textbf{B}^{ac}$ is a vector with unitary length indicating the direction of the oscillating B field of the microwave. The $\bra{e1\downarrow}$ and $\ket{e2\uparrow}$ are the two spin states of the SnV under external, static B field. The Ham reduction factor $q$ of the ground state is set to 0.471 according to \cite{thiering2018ab}. First we investigate the effect of strain magnitude to the transverse and longitudinal $g$ factor. The result is shown in Figure \ref{subfig:MW g factor} (a), indicating a $g$ factor of 1.64. Here the static \textbf{B} field set to be \SI{80}{\milli\tesla} along the quantization axis, which is in line with our experimental setup. We then investigate the angular dependence of the transverse $g$ factor with different static field orientations using our experimentally observed strain profile. We note that the $g$ factor has a weak angular dependence, indicating a consistently efficient MW driving efficiency regardless of the static B field orientation, highlighting the robustness of the strained SnV centers.

\subsection{Device info}
\begin{figure}[ht]
\centering
\includegraphics[width=0.7\textwidth]{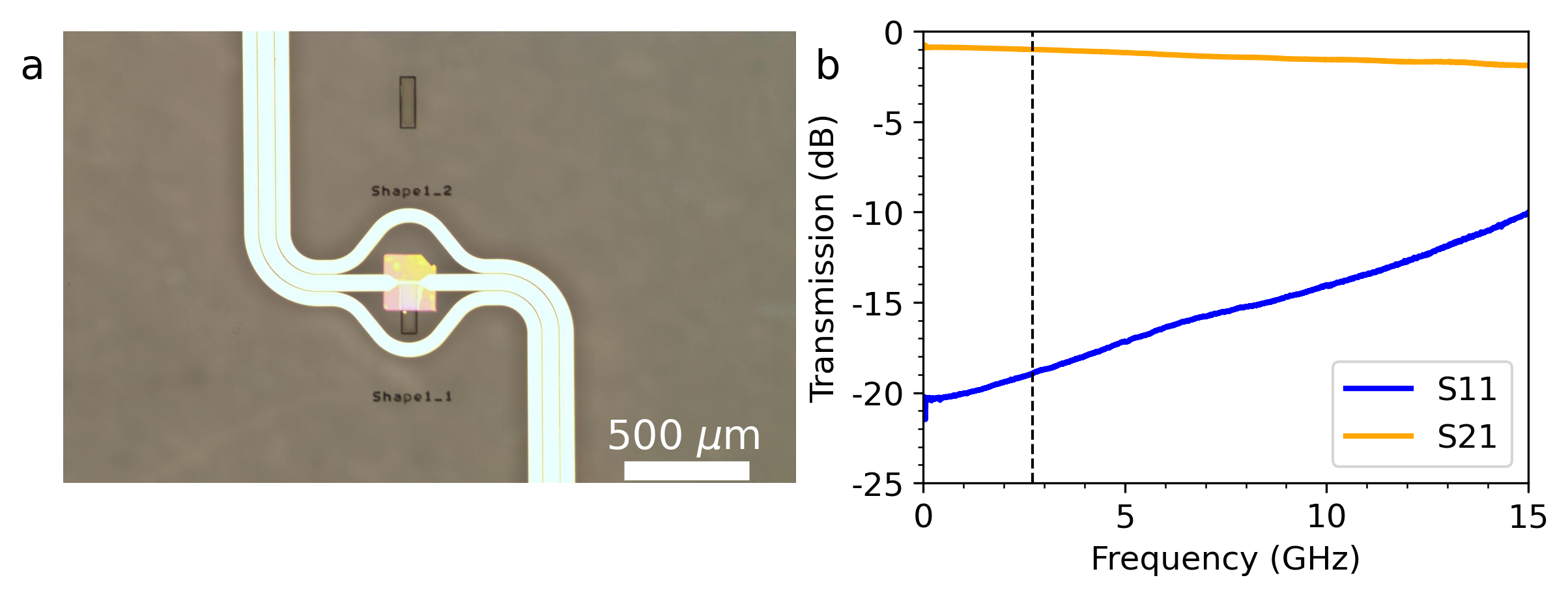}
\caption{Microwave CPW. (a) Microscope image of the CPW. The bonding pad for wire bonding is not shown. (b) The transmission of the CPW measured via a probe station using a vector network analyzer (VNA). The S11 is the reflection spectrum while the S21 is the transmission. The operation frequency (\SI{2.75}{\giga\hertz}) is labelled as a black, dashed line.}
\label{subfig:MW device}
\end{figure}

In this work we utilized on-chip coplanar waveguide (CPW) to deliver microwave signals to target SnV centers. Compared with wire-bonded metal striplines \cite{nv_microwave_wire}, lithography-defined CPW offers deterministic and reproducible microwave power and magnetic field strength at target location. We designed our waveguide to match the impedance ($50 \Omega$) of other electronics in the setup. Ignoring the local dielectric variation near the diamond membrane region, we designed the layout of the CPW based on the permittivity of the fused silica (3.7). The width of the center and the gap is set to \SI{60}{\micro\meter} and \SI{6}{\micro\meter}, respectively. To enhance the local field strength near the SnV region on the membrane, the center of the CPW is reduced to \SI{6}{\micro\meter}. The ground lines of the waveguide is designed to across the membrane to compensate for the trench design, offering a balanced microwave delivering mode. We used a two-port microwave transmission design, demonstrating the potential of driving centers in multiple on-chip devices in the future. The two-port design also allows the microwave signal to be transmitted and dissipated outside of the chamber, relaxing the requirements for reflected signal management (such as circulator). The microscope image of the whole CPW design is shown in Figure \ref{subfig:MW device} (a), with the transmission data of an identical device shown in Figure \ref{subfig:MW device} (b). We show that the transmission loss is low from dc to \SI{15}{\giga\hertz}, with the thermal loss ($P_{\text{in}} - P_{\text{out}} - P_{\text{reflected}}$) around the operation point (\SI{2.75}{\giga\hertz}) to be 1.5 dB. 

\subsection{MW field simulation}
\begin{figure}[ht]
\centering
\includegraphics[width=0.8\textwidth]{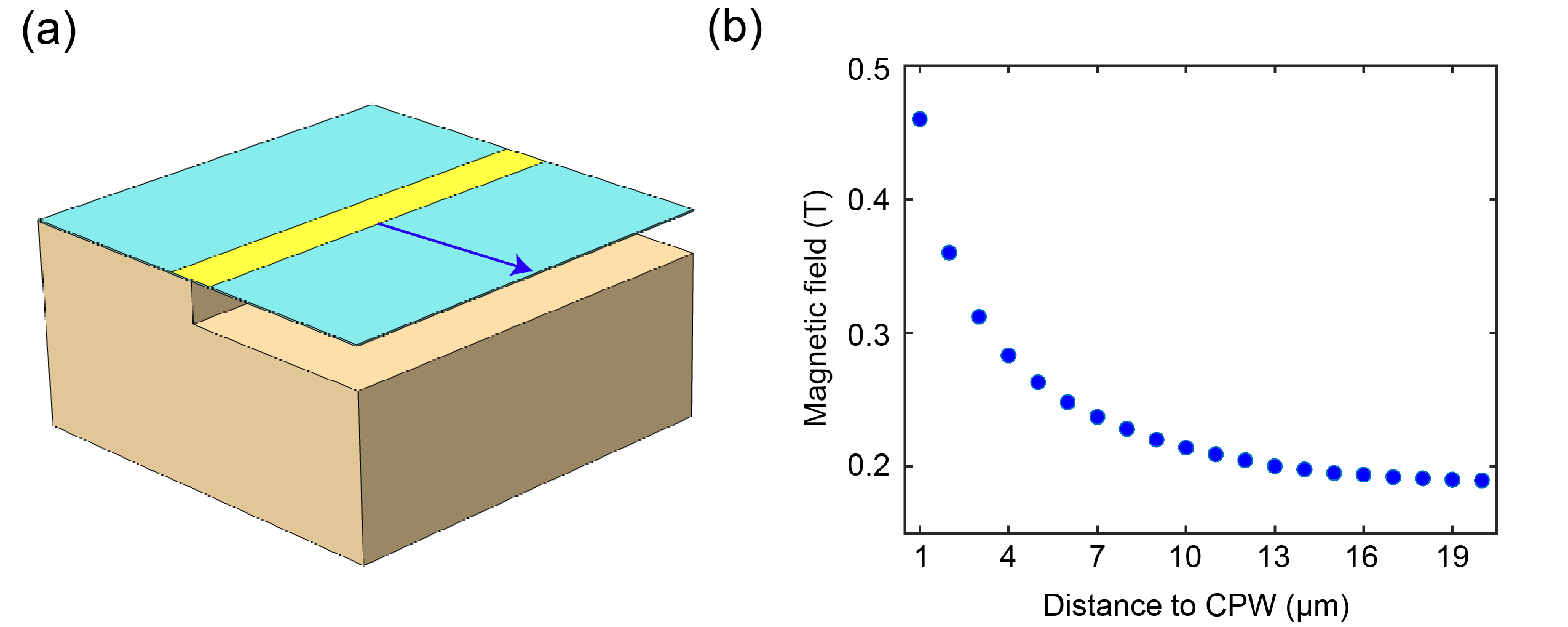}
\caption{Simulation of the microwave field in diamond membrane. (a) The 3D structure of the simulated device. The length (width) of the suspended diamond membrane is set to \SI{50}{\micro\meter} (\SI{25}{\micro\meter}). (b) The magnetic field in diamond membrane, as a function of distance to the coplanar waveguide. The simulated depth is \SI{40}{\nano\meter} from the top surface of diamond, and the simulated position is labeled as a blue arrow in (a).}
\label{fig:mwsimulation}
\end{figure}
We use COMSOL to simulate the magnetic field acting on nearby color centers. In the simulation, we set the microwave drive power to 24 dBm, the microwave drive frequency to \SI{2.75}{\giga\hertz}, and set the characteristic impedance of the coplanar waveguide to 50 Ohm. The simulated structure and the magnetic field distribution are presented in \autoref{fig:mwsimulation}. Based on the simulation results, we expect the effective B field applied to the color centers ranges from \SI{0.2}{\tesla} to \SI{0.4}{\tesla}, corresponding to a transverse B field of \SI{0.12}{\tesla} to \SI{0.23}{\tesla}.

\subsection{Heating effect of the system}
To investigate the power dependence of the SnV Rabi oscillation, we sweep the MW drive power and extract the Rabi frequency. We observe the expected $\sqrt{p}$-behaviour for low drive powers $p<24$ dBm, but a clear deviation for larger drive powers. All the power and voltages are referred to the estimated value on the device, extracted by a separate calibration measurements in transmission geometry. We note that no increase in cryostat temperature is observed during the pulsed Rabi measurement.

\begin{figure}[ht]
\centering
\includegraphics[width=0.35\textwidth]{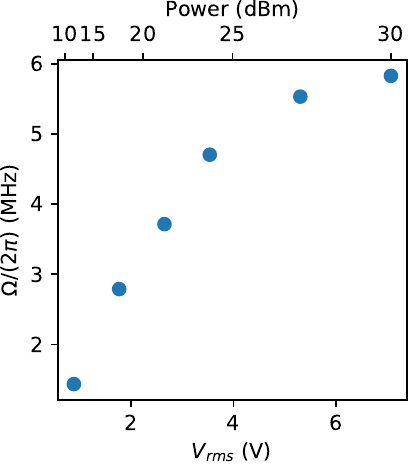}
\caption{Fitted Rabi-frequency $\Omega/2 \pi$ over MW amplitude and power, showing the expected $\sqrt{p}$-behaviour for low drive powers $p<24$ dBm. }
\label{subfig:PowerRabi}
\end{figure}

\begin{figure}[ht]
\centering
\includegraphics[width=\textwidth]{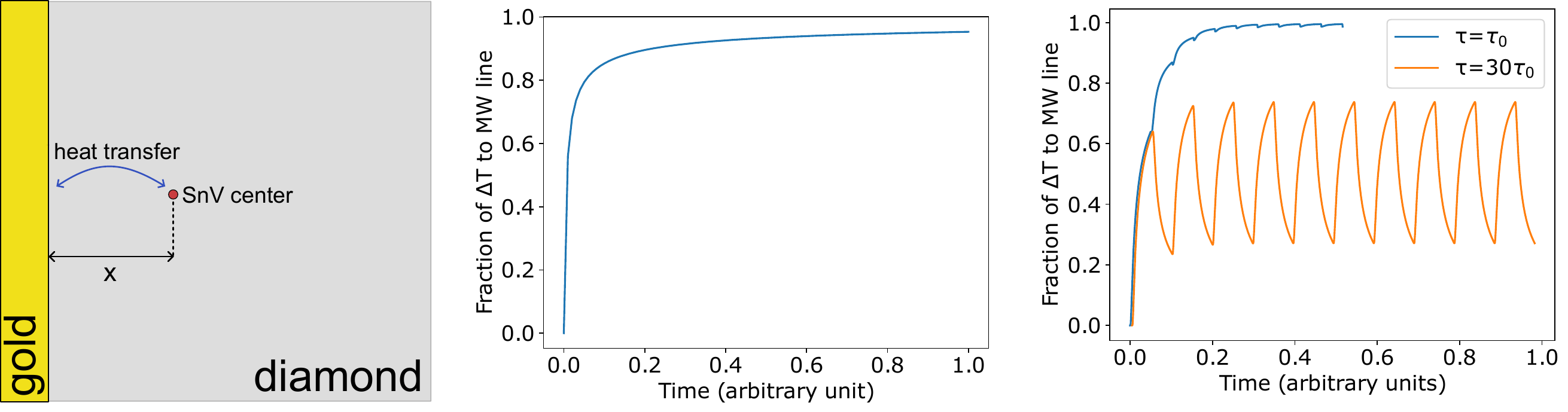}
\caption{(a) The SnV center as a point like object at a fixed distance x from the microwave gold transmission line. This is the model underlying the 1D heat equation solution. (b) Exemplary temperature increase vs time at the emitter position due to continuous microwave drive. The value is expressed as a fraction of the initial temperature difference approaching equilibrium. (c) Exemplary pulse sequence applying 10 successive microwave pulses each with intermediate spacing of $\tau_0$ and $30 \tau_0$, respectively. The maximum temperature differs significantly.}
\label{subfig:MW heat}
\end{figure}

The effect of heating on the emitter can be modeled as depicted in figure \ref{subfig:MW heat} (a) where we follow the approach taken in Ref \cite{lukin_heating}. The emitter is treated as a point-like object at a fixed distance $X$ from the microwave line. Compared to the SnV center the extent of the gold strip is well approximated as infinite such that we can model this as a 1D problem. Assuming the gold heats and cools instantaneously at the beginning and end of a microwave pulse, a solution to the 1D heat equation yields a temperature increase at the SnV center $\Delta T_{SnV} \propto erf\left(\frac{d}{\sqrt{4 \alpha t}}\right)$ where $\alpha$ is the thermal diffusivity in diamond. Figure \ref{subfig:MW heat} (b) shows that the temperature at the emitter at asymptotically approaches the temperature of the microwave line. When higher Rabi frequencies are used another interesting effect is observed for more complex pulse sequences. Figure \ref{subfig:MW heat} (c) shows the effect of a sequence of pulses with a constant interpulse spacing $\tau_0$. If $\tau_0$ is significantly smaller than the time per pulse, the heat cannot flow away fast enough such that a net heating effect is observed per applied pulse. This means at high Rabi frequencies the coherence time of the spin can depend on the time between pulses. \\

\clearpage

\section{Additional optical properties of the strained SnV}
\subsection{Polarization of the SnV}
\begin{figure}[ht]
\begin{subfigure}{0.45\textwidth}
\centering
\includegraphics[width=\textwidth]{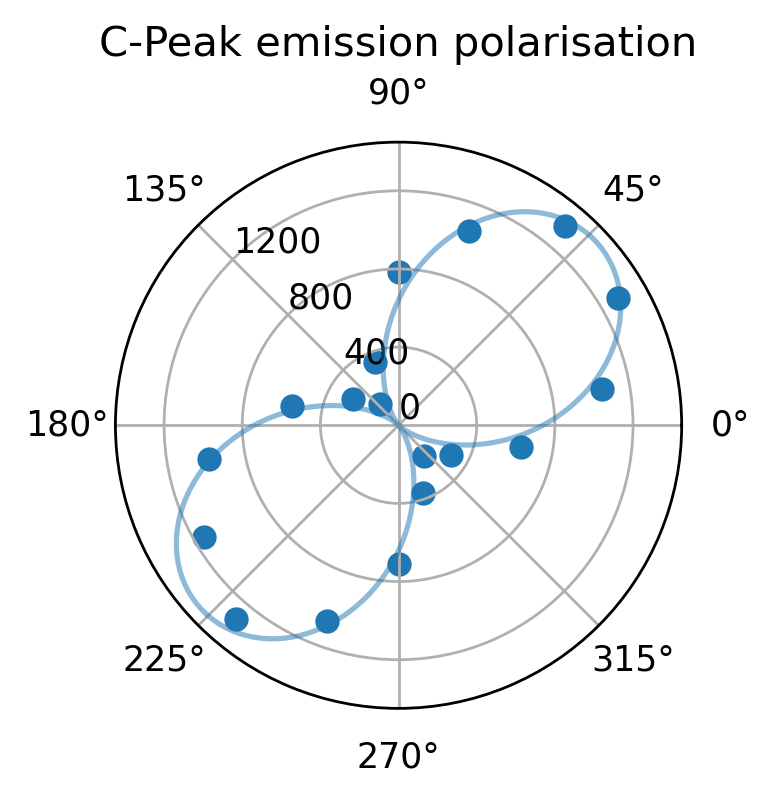}
\caption{}
\label{subfig:CPeak}
\end{subfigure}
\hfill
\begin{subfigure}{0.45\textwidth}
\centering
\includegraphics[width=\textwidth]{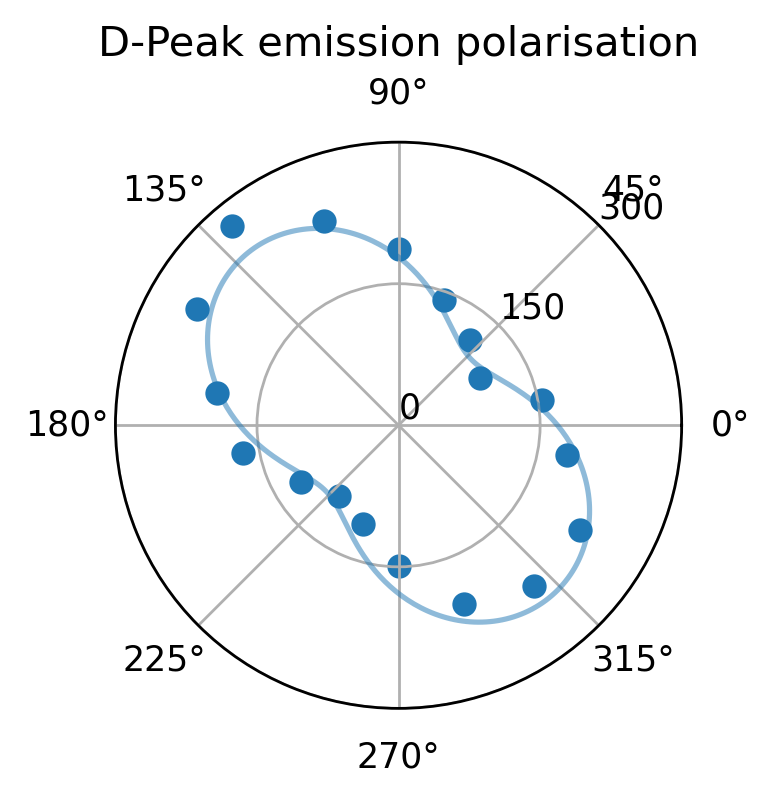}
\caption{}
\label{subfig:DPeak}
\end{subfigure}
\caption{Intensity of the (a) C-peak and (b) D-peak PL emission over linear polarisation in the lab-frame.}
\end{figure}

We probe the polarization of the strained SnVs by inserting a motor-mounted half-wave-plate and a linear polarizer in the detection path. 
The total intensity of the C-peak and the D-peak over polarisation angle are shown in Fig. \ref{subfig:CPeak}) and Fig. \ref{subfig:DPeak}), respectively. No magnetic field was applied.  $0^\circ$ in the graph indicates the magnet $x$-axis. The solid line is a fit of the expected polarisation, linear for the C-peak and circular for the D-peak, projected into the lab-frame according to the model in Ref. \cite{Hepp2014}. Both figures indicate a polarisation behaviour commensurate with bulk group IV color centers, showing that the polarisation is not changed when introducing strain. 

\subsection{Optical lifetime}
\begin{figure}[ht]
\centering
\includegraphics[width=0.6\textwidth]{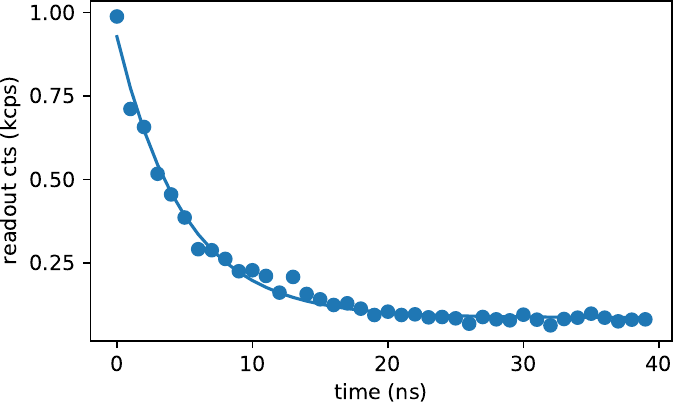}
\caption{Optical lifetime measurement of the C-transition. The solid line is a fit by a single-exponential, showing a decay time of \SI{4.93\pm0.19}{\nano\second}.}
\label{subfig:Lifetime}
\end{figure}

We extract the optical lifetime of the SnV by driving the C-transition at zero magnetic field with a single EOM-sideband and turning it off abruptly. The fall time is limited to \SI{200}{\pico\second} by the EOM. The decay time of the single-exponential is \SI{4.933\pm0.190}{\nano\second} which is similar to the bulk value, as shown in Fig. \ref{subfig:Lifetime} \cite{trusheim2020transform}.

\subsection{Power saturation}
We extract the initialization rate, optical cyclicity and saturation power by prior knowledge of the optical lifetime and by sweeping the laser power \cite{Debroux2021}. The initialisation rates are fitted by $\frac{1}{\eta}\frac{\Gamma}{2}\frac{p/p_\text{sat}}{1+p/p_\text{sat}}$ and we extract a saturation power of \SI{7.96}{\nano\watt} and an optical cyclicity of $\eta \approx 2018$. For the microwave spin control measurement we operate at a saturation parameter of $s=p/p_\text{sat}\approx 10$ for the initialization and readout pulses.

\begin{figure}[ht]
\centering
\includegraphics[width=0.6\textwidth]{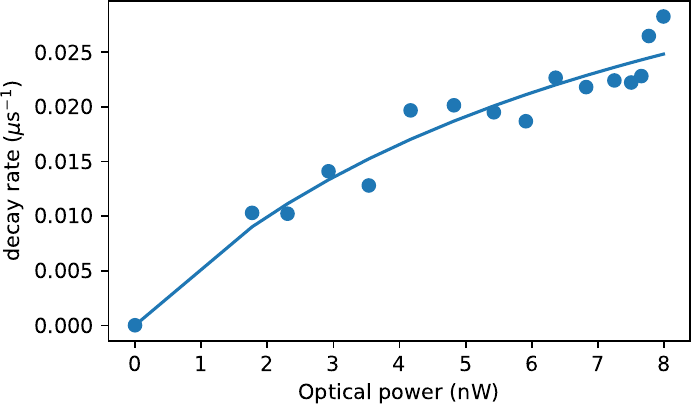}
\caption{Initialization rates for different laser powers with a fit by $\frac{1}{\eta}\frac{\Gamma}{2}\frac{p/p_\text{sat}}{1+p/p_\text{sat}}$}
\label{subfig:SatCurve}
\end{figure}

\subsection{Long term stability of PLE}
\begin{figure}[ht]
\centering
\includegraphics[width=0.55\textwidth]{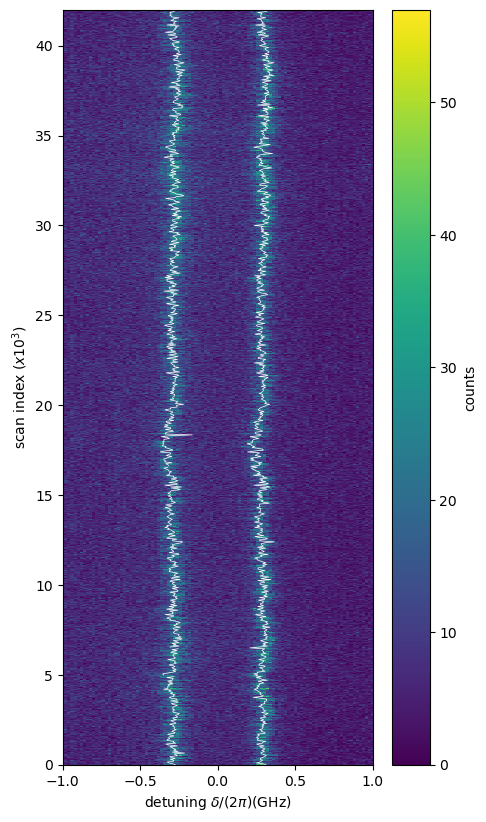}
\caption{Time evolution of the PLE line shape. Each vertical cut is the average of fast PLE scans over \SI{1}{\second}, with each shot of the measurement containing both green and red excitation. The solid white line is the fitted emitter resonance frequency.}
\label{subfig:LongtimePLE}
\end{figure}

We acquire PLE for more than 11 hours to test the long-time stability of the SnV (see Fig. \ref{subfig:LongtimePLE}). We observe a very good frequency stability and only modest spectral wandering. 
\clearpage

\begin{figure}[H]
\centering
\begin{subfigure}{0.49\textwidth}
\centering
\includegraphics[width=\textwidth]{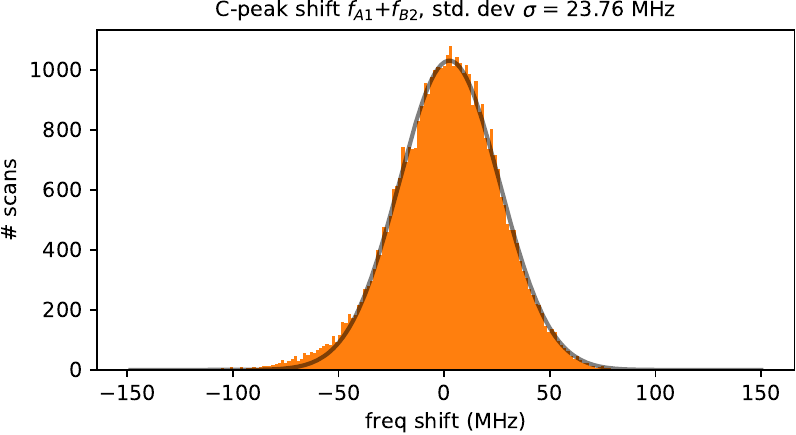}
\caption{}
\label{subfig:CommonMode}
\end{subfigure}
\hfill
\begin{subfigure}{0.49\textwidth}
\centering
\includegraphics[width=\textwidth]{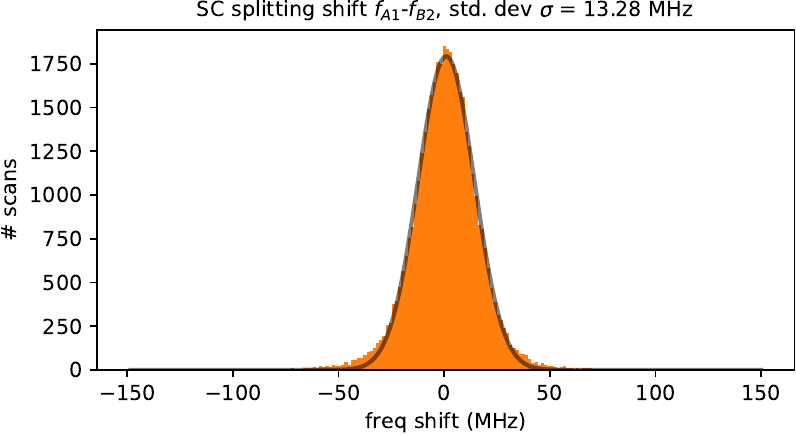}
\caption{}
\label{subfig:SplittingShift}
\end{subfigure}
\caption{(a) Histogram of the mean C-peak frequency from shot-to-shot. (b) Histogram of the spin-conserving splitting from shot-to-shot. The grey line is a fit with a Gaussian distribution. }
\end{figure}

We fit each acquired PLE trace and extract the common mode shift of the spin-conserving transition \cite{arjona_indistinguishability} (Fig. \ref{subfig:CommonMode}).The Gaussian distribution of the shot-to-shot center frequencies has a standard deviation of $\sigma=$\SI{23.8 \pm 0.1}{\mega\hertz}. Similarly, the distribution of extracted spin-conserving splittings (Fig. \ref{subfig:SplittingShift}) has a standard deviation of only $\sigma=$\SI{13.28 \pm 0.06}{\mega\hertz}.

\section{Optical control of the SnV spin}
\subsection{Optical splitting with external B field} \label{section:optical splitting with B field}
We scan the magnetic field over the whole sphere at fixed magnitude. The path between the approximately equidistant points is numerically minimised. The hysteresis of the $B$-field is on the order of \SI{10}{\percent} as estimated from linear sweeps along a single magnet axis.

\begin{figure}[ht]
\centering
\includegraphics[width=0.9\textwidth]{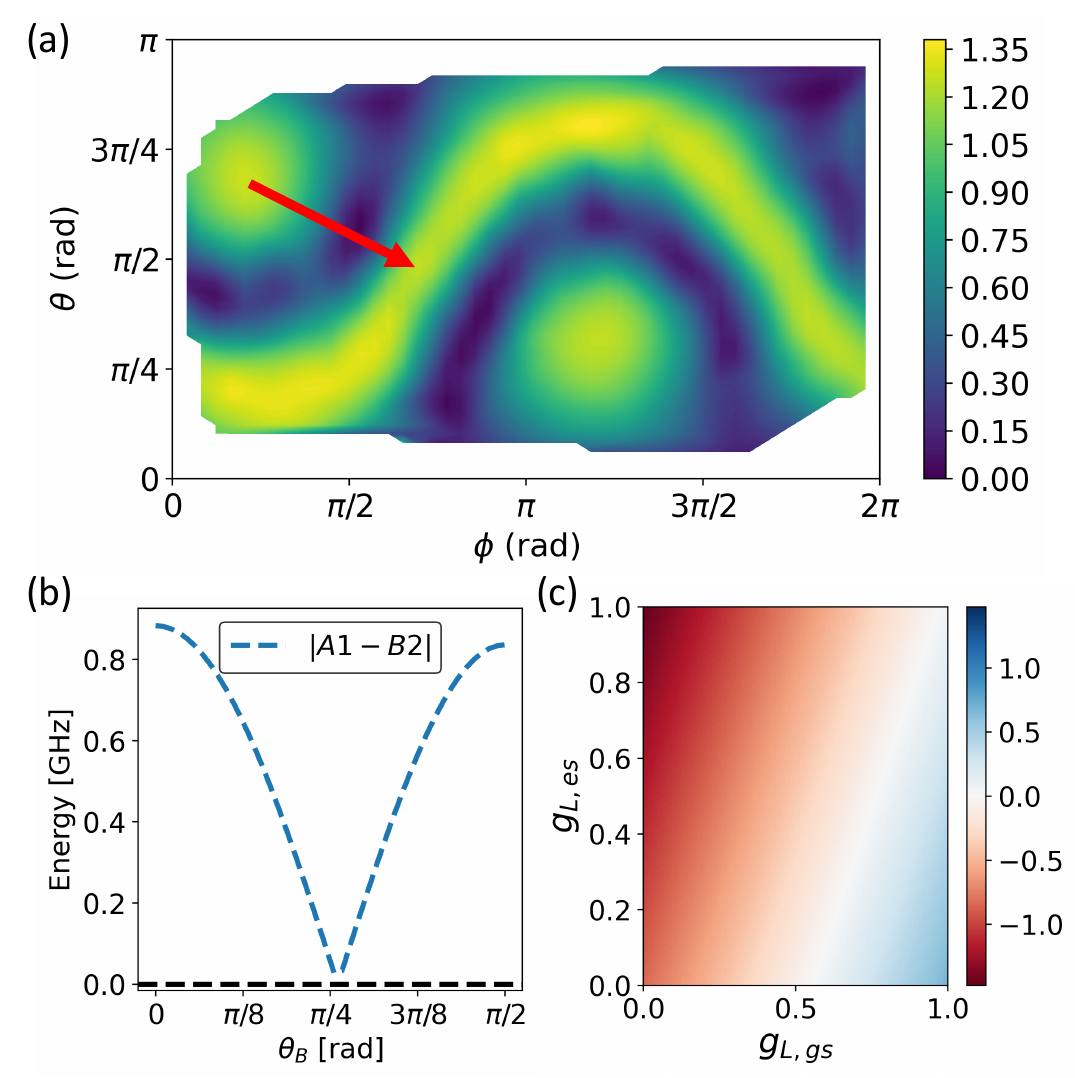}
\caption{Experimental measured and simulations of the spin-conserving optical transitions $A1, B2$ with varying external magnetic field $B$. (a) Scanning of the splittings of $A1, B2$ transitions with varying $B$ field directions. The magnitude of $B$ is set to \SI{0.2}{\tesla}. The $x, y$ axis represents the azimuthal ($\phi$) and polar angle ($\theta$) of the $B$ field in the Lab frame. The two poles on the plot represent directions along the quantization axis of the \ch{SnV-} and the belt represents the equator. (b) Simulated splittings of the $A1, B2$ transitions by diagonalizing the system Hamiltonian along a chosen path of varying $B$ fields, where the path is depicted as a red arrow in (a). The $x$ axis represents the polar angle of the $B$ field in the defect frame. Simulation agrees qualitatively with experiments with the magnitude of splitting underestimated by \SI{0.4}{\giga\hertz}. (c) The differences between the $A1, B2$ splittings at $\theta_B = 0$ and $\theta_B = \pi/2$ with varying Steven's reduction factor. The white region corresponds to pairs of Steven's reduction factor for ground and excited states, when taken into the diagonalized Hamiltonian, that match the experimental observations.}
\label{subfig:optical_splitting_B_field}
\end{figure}

The splitting of the $A1, B2$ optical transitions with varying $B$ fields can be computed by diagonalizing the system Hamiltonian $H_{\text{sys}}$ of Eq.~\ref{eq:system_hamiltonian}, and the results are shown in Fig~\ref{subfig:optical_splitting_B_field} (b) and (c). When constructing the Hamiltonian, we considered the Steven's term $g_L$ in the reduction factor $q$ as a free parameter. The Steven's term, as discussed in subsection~\ref{subsection: Zeeman}, originates from the defect symmetry being lower than $O(3)$. Here we determined the range of $g_L$ by matching the experiments. We plot the difference of the splitting when the $B$ field is aligned with the defect quantization axis ($\theta_B = 0$), and aligned along the equator ($\theta_B = \pi/2$) with varying $g_L \in [0, 1]$ in Fig~\ref{subfig:optical_splitting_B_field} (c). The white region in the plot (values close to zero) corresponds to the two splittings being close in energy, matching the experimental observations. Therefore our calculations enabled the narrowing down of the the possible values of Steven's factor to $g_{L, \text{gs}} \in [0.5, 1.0]$ and $g_{L, \text{es}} \sim 2g_{L, \text{gs}} - 1$.

\clearpage


\subsection{Optical cyclicity of the SnV}
\begin{figure}[H]
\begin{subfigure}{0.45\textwidth}
\centering
\includegraphics[width=\textwidth]{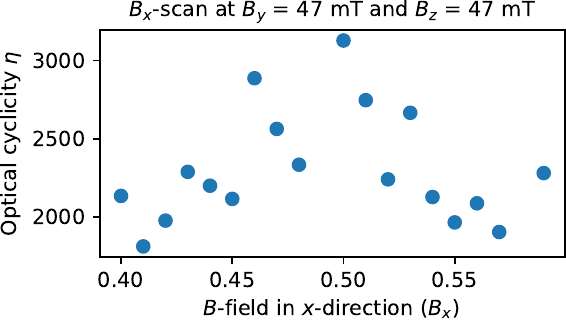}
\caption{}
\label{subfig:Bx_scan}
\end{subfigure}
\hfill
\begin{subfigure}{0.45\textwidth}
\centering
\includegraphics[width=\textwidth]{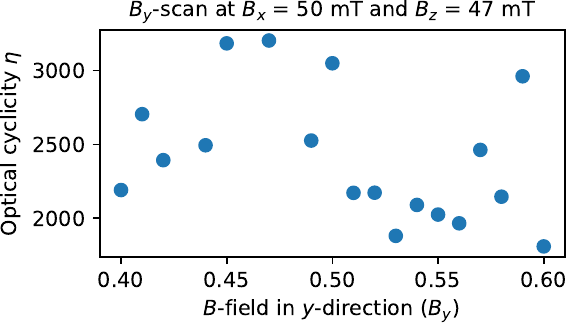}
\caption{}
\label{subfig:By_scan}
\end{subfigure}
\caption{Optical cyclicity at (a) fixed $B_y$ and $B_z$, with swept $B_x$ and (b) fixed $B_x$ and $B_z$, with swept $B_y$.}
\end{figure}

We coarsely align the $B$-field by matching it to the polarisation of the optical dipoles (see SI section 3.1) and obtain an optical cyclicity of $\eta \approx 2018$. The cyclicity has a single local maximum close to the pole of the emitter axis, such that we can increase it by sweeping two of the three magnet axes independently. We extract the cyclicity by measuring the decay rate of one of the spin-conserving transitions. The frequency of the sidebands driving the transitions is fixed, noting that the change in $B$-field magnitude corresponds to a change in spin-conserving splitting within one optical linewidth. One can see only a modest increase in cyclicity in Fig. \ref{subfig:Bx_scan} and \ref{subfig:By_scan}, such that we conclude that strain limits the maximum achievable cyclicity. Nevertheless, the error introduced by the finite cyclicity will be negligible in spin-photon entanglement protocols due to the relatively high value and enable single-shot readout with nanostructures or microcavities.

The optical cyclicity, as a branching ratio between spin-conserving and spin-flipping transitions, depends on both system properties and the external optical field. Theoretically, we can only investigate the system property side. Here we investigate an alternative problem---spontaneous emission rate ratio between spin-conserving and spin-flipping transitions---optical cyclicity with absence of the external optical excitation.

\begin{figure}[ht]
\centering
\includegraphics[width=0.6\textwidth]{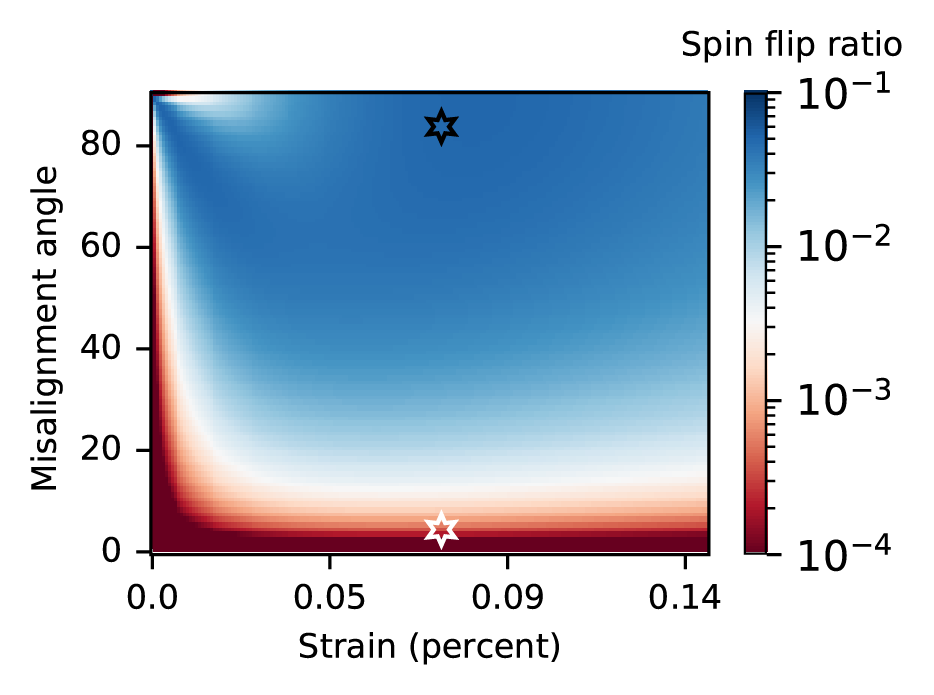}
\caption{The calculated cyclicity of the SnV with respect to the $B$ field polar angle $\theta$ and the strain magnitude. The operation point of the MW-based (all-optical) control of the SnV spin qubit is highlighted in white (black) stars.}
\label{subfig:cyclicity}
\end{figure}

According to Ref~\cite{Hepp2014}, we use optical dipole matrices to calculate the emission rate of the two transitions. The rate (probablity) can be expressed using Fermi's Golden Rule:
\begin{equation}
    P =2 \pi \rho |\bra{\psi_f}|e \cdot \hat{\textbf{r}} |\ket{\psi_i}|^2 = 2 \pi \rho |\bra{\psi_f}|\hat{\textbf{p}} |\ket{\psi_i}|^2
\end{equation}
Where $\rho$ is the density of states where we set to 1, the $\ket{\psi_f}$ and $\ket{\psi_i}$ are the final and initial state of the SnV which we assign to the excited state minimum $\ket{e_A \downarrow}$ and ground states $\ket{e_1 \downarrow}$ ($\ket{e_2 \uparrow}$) for spin conserving (flipping) transitions. The transition probablity (rate) $P$ is related to the optical dipole $\hat{\textbf{p}}$ which is defined as:

\begin{align}
    \hat{p_x} = e\left[\begin{matrix}
    1 & 0 & 0 & 0\\
    0 & 1 & 0 & 0\\
    0 & 0 & -1 & 0\\
    0 & 0 & 0 & -1
    \end{matrix}\right],
    \hat{p_y} = e\left[\begin{matrix}
    0 & 0 & -1 & 0\\
    0 & 0 & 0 & -1\\
    -1 & 0 & 0 & 0\\
    0 & -1 & 0 & 0
    \end{matrix}\right],
    \hat{p_z} = e\left[\begin{matrix}
    1 & 0 & 0 & 0\\
    0 & 1 & 0 & 0\\
    0 & 0 & 1 & 0\\
    0 & 0 & 0 & 1
\end{matrix}\right]
\end{align}
Using the above definition, we can calculate the spin flip ratio which is the inverse of spontaneous cyclicity $\frac{1}{\eta}=\frac{P_{\text{flipping}}}{P_{\text{conserving}}}$ with respect to the strain magnitude and the $B$ field polar angle $\theta$, as shown in Figure \ref{subfig:cyclicity}. The operation point of the MW-based (all-optical) control of the SnV spin qubit is highlighted in white (black) stars, showing a cyclicity of \SI{\geq2000}{} if $\theta<\SI{4}{\degree}$ and a cyclicity of \SI{\approx20}{} if $\theta>\SI{85}{\degree}$, in a rough agreement with the experimentally observed values. We note that the presence of the moderate-level strain will make the overall cyclicity lower than the strain-free case, but the achievable value is still compatible with single shot readout requirements if the signal count can be improved by device design or setup optimization. 

\subsection{All-optical spin control and operation at perpendicular $B$-fields}
\begin{figure}[ht]
\centering
\includegraphics[width=0.5\textwidth]{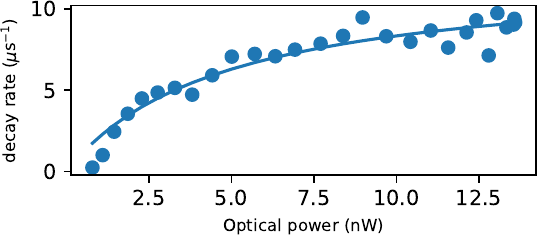}
\caption{Initialization rates for different laser powers with a fit by $\frac{1}{\eta}\frac{\Gamma}{2}\frac{p/p_\text{sat}}{1+p/p_\text{sat}}$}
\label{subfig:SatCurvePerp}
\end{figure}

We implemented the all-optical control technique shown in Ref. \cite{Debroux2021} on Device 1 on a strained SnV with a ground state splitting of $\Delta_\text{GS}=$\SI{1384}{\giga\hertz}. We extracted an optical lifetime of \SI{7.04\pm0.10}{\nano\second} which is compatible for an SnV in proximity of a surface \cite{gorlitz2020}. We set the magnetic field to $|B| = 100 $ mT perpendicular to the emitter axis. From the saturation power measurement in Fig. \ref{subfig:SatCurvePerp} we extract a saturation power of $p_\text{sat} = $\SI{4.82\pm0.81}{\nano\watt} and a cyclicity of $\eta = $ \SI{5.78\pm0.36}. The low saturation power and low cyclicity indicate that efficient all-optical control should be possible in principle. 

\begin{figure}[ht]
\centering
\includegraphics[width=0.8\textwidth]{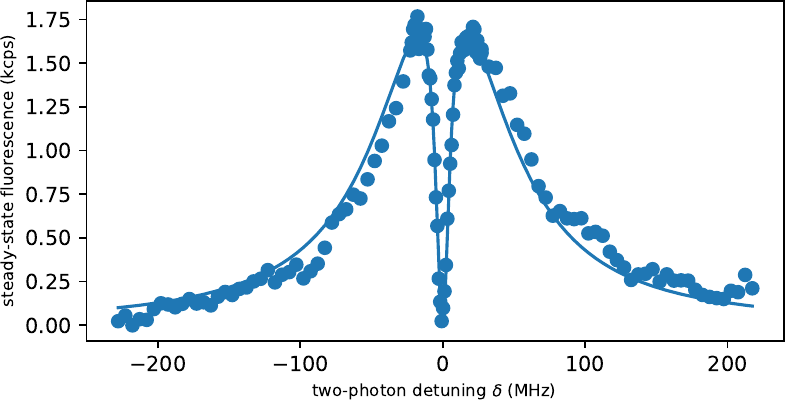}
\caption{Coherent population trapping measured on an SnV at perpendicular magnetic fields.}
\label{subfig:CPT}
\end{figure}

We implement an optical lambda system on the spin-conserving transition A1 and spin-flipping transition A2 and measure coherent population trapping (CPT) by driving both transitions simultaneously (Fig. \ref{subfig:CPT}). Fitting the data with the model in Ref. \cite{Debroux2021,fleischhauer2005electromagnetically}, we get an excited state decay rate of  $\Gamma/2 \pi = $\SI{26.52\pm0.91}{\mega\hertz}, only a factor of 1.17 larger than the transform-limited linewidth $\Gamma_0/(2 \pi)$ = \SI{22.60\pm0.05}{\mega\hertz}.

\begin{figure}[ht]
\centering
\includegraphics[width=0.7\textwidth]{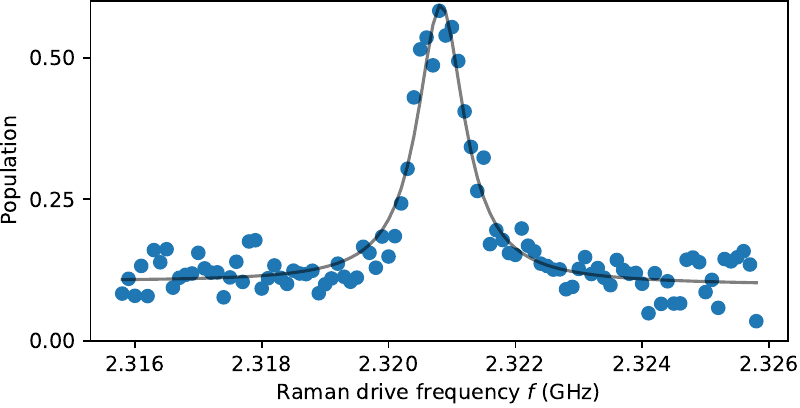}
\caption{All-optical pulsed ODMR at perpendicular $B$-field directions.}
\label{subfig:ODMRAllOptical}
\end{figure}

For all-optical Raman control, we choose to operate at a single-photon detuning of $\Delta = 1.5$ GHz. At the lowest laser sideband powers of $p=455$ nW, we get an intrinsic ODMR linewidth of $\delta f =1/T_{2^*} = $ \SI{899\pm54}{\kilo\hertz} (Fig. \ref{subfig:ODMRAllOptical}). The qubit frequency of $f_\text{qubit} = 2.321 $ GHz, yields a $g$-factor of $g=0.83$, meaning that perpendicular fields can couple to the SnV efficiently due to strain.

\begin{figure}[ht]
\centering
\includegraphics[width=0.6\textwidth]{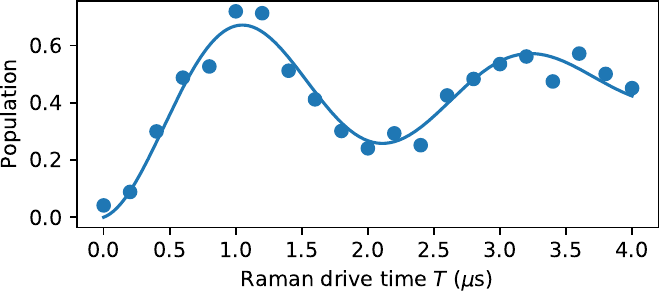}
\caption{All-optical spin control at perpendicular $B$-field directions and sideband power of $p=1012$ nW.}
\label{subfig:RabiAllOptical}
\end{figure}

We sweep the Raman drive time $T$ at laser sideband powers of $p=1012$ nW and extract a Rabi frequency of $\Omega/2 \pi$ = \SI{450\pm47}{\kilo\hertz}(Fig. \ref{subfig:RabiAllOptical}) and a $\pi$-gate fidelity of $F_\pi = $ \SI{83\pm2}{\percent}, similar to Ref. \cite{Debroux2021}. 

\begin{figure}[ht]
\centering
\includegraphics[width=0.6\textwidth]{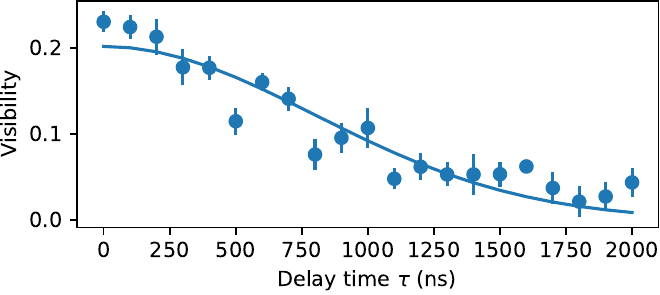}
\caption{All-optical Ramsey measurement showing an inhomogeneous dephasing time of $T_{2*} = 1.13 \pm 0.07$ µs}
\label{subfig:RamseyAllOptical}
\end{figure}

\begin{figure}[ht]
\centering
\includegraphics[width=0.6\textwidth]{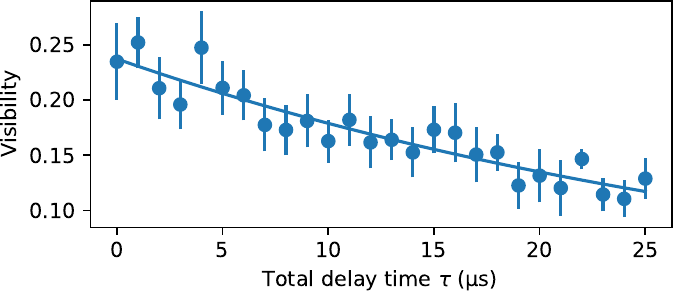}
\caption{All-optical Hahn-Echo measurement showing a dephasing time of $T_{2} = 35.45 \pm 2.96 $ µs}
\label{subfig:HahnEchoAllOptical}
\end{figure}

Ramsey measurements (Fig. \ref{subfig:RamseyAllOptical}) yield an inhomogeneous dephasing time of $T_{2*} = $ \SI{1.13\pm0.07}{\micro\second} and a Hahn-Echo measurements (Fig. \ref{subfig:HahnEchoAllOptical}) a dephasing time of $T_{2} = $ \SI{35.5\pm3.0}{\micro\second}.

\begin{figure}[ht]
\centering
\includegraphics[width=0.6\textwidth]{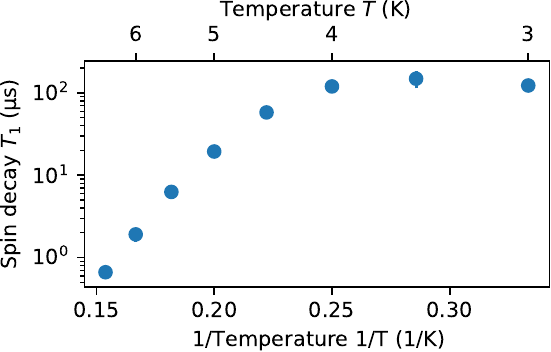}
\caption{Spin decay $T_1$ over inverse temperature $1/T$. }
\label{subfig:SpinT1}
\end{figure}

We additionally measured the spin decay time $T_1$ at the perpendicular field orientation and found much shorter times on the order of 100 µs (Fig. \ref{subfig:SpinT1}).

\FloatBarrier

\subsection{Spin $T_1$ analysis}
Phonon-induced depolarization of group IV centers' spin is the dominant source of decoherence. Therefore, improving the spin decay time $T_1$ is the central task to enhance the coherence of the system. As discussed in Ref. \cite{meesala2018strain}, the spin decay time $T_1$ in the group IV has two meanings, the orbital relaxation time $T_1^{\text{orbit}}$ between the energy levels in different orbital branches but with same spin projection, and the $T_1^{\text{spin}}$ between the qubit levels with frequency $\omega_s$. Ultimately, the $T_1^{\text{spin}}$ is the factor that directly relates to the coherence of the SnV, so we limit our $T_1$ discussion to $T_1^{\text{spin}}$ only. 

There are three phonon-induced $T_1^{\text{spin}}$ decay path \cite{meesala2018strain}, including direct single phonon, resonant two phonon (Orbach process) \cite{Orbach1961}, and off-resonant two phonon (Raman process). Similar to SiV, the SnV shows much slower single phonon and Raman spin decay, so we focus on the Orbach process and study its dependence with the $B$ field orientation ($\theta$) and the strain magnitude. Adapted from Ref. \cite{Orbach1961}, we write the decay rate $\gamma_{spin}^{2}$ as follows:
\begin{equation}
\gamma_{spin}^{2} \propto \frac{ \Delta_{gs}^3}{\exp{(h\Delta_{gs}/{k_BT})}-1} \frac{\left|\sum \bra{e_1\downarrow}H_\epsilon^{AC}\ket{e_j}\bra{e_j}H_\epsilon^{AC}\ket{e_2\uparrow}\right|^2}{ \sum\left|\bra{e_i}H_\epsilon^{ac}\ket{e_j}\right|^2}
\end{equation}
where $i$ represent the states of the lower orbital branch ($\ket{e_1\downarrow}$, $\ket{e_2\uparrow}$) and $j$ represent the states of the upper orbital branch ($\ket{e_3\downarrow}$, $\ket{e_4\uparrow}$). The $H_\epsilon^{ac}$ denotes an AC strain field which correlates to the phonon interaction in the crystal. We used balanced magnitude for the $H_{\epsilon_{E_x}}^{AC}$ and $H_{\epsilon_{E_y}}^{AC}$ by setting the $H_\epsilon^{ac}$ as follows:
\begin{align}
    H_\epsilon^{ac} = e\left[\begin{matrix}
    -1 & 0 & 1 & 0\\
    0 & -1 & 0 & 1\\
    1 & 0 & 1 & 0\\
    0 & 1 & 0 & 1
    \end{matrix}\right],
\end{align}

\begin{figure}[ht]
\centering
\includegraphics[width=0.5\textwidth]{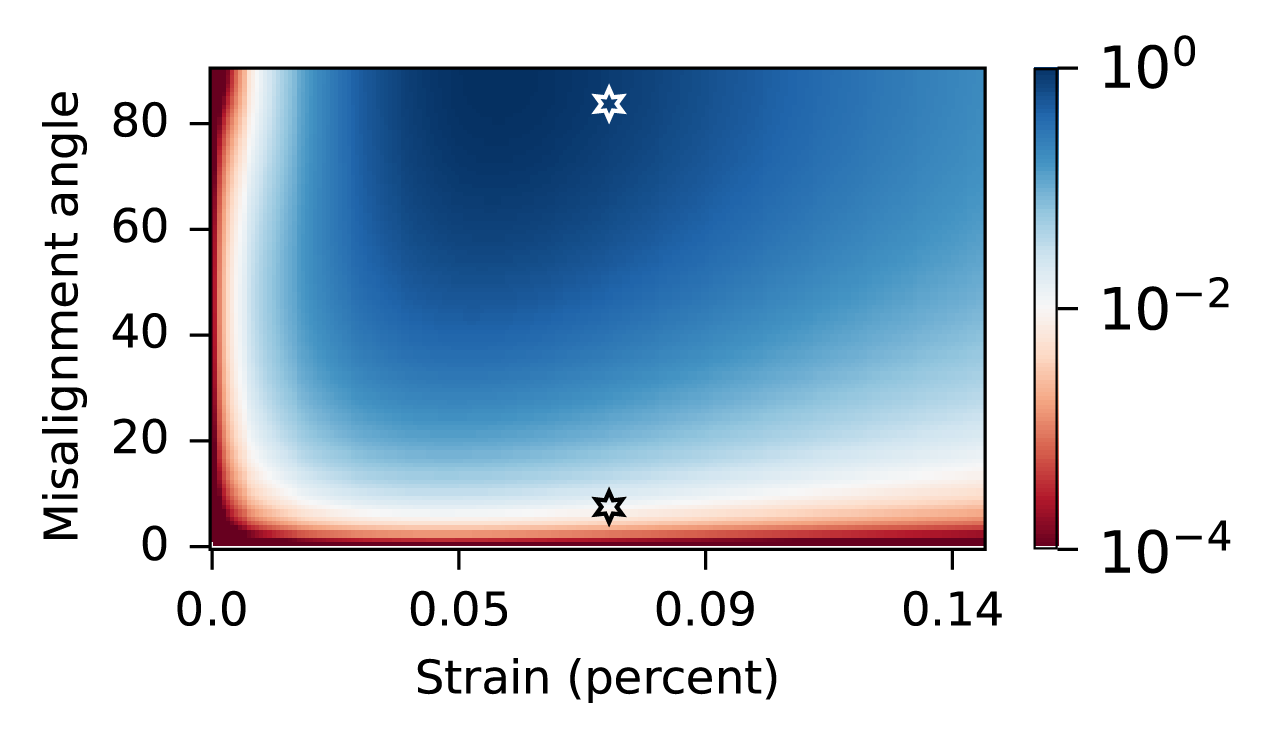}
\caption{The calculated Orbach decay rate of the SnV with respect to the $B$ field polar angle $\theta$ and the strain magnitude. The operation point of the MW-based (all-optical) control of the SnV spin qubit is highlighted in black (white) stars.}
\label{subfig:spin T1}
\end{figure}

The relative decay rate at temperature \SI{4}{\kelvin} with the maximum normalized to 1 is shown in Figure \ref{subfig:spin T1}, with MW-based (small $\theta$) and all-optical (large $\theta$) operation points for spin control of the SnV highlighted in black (white) stars. We  observe a ratio of \SIrange{500}{1200}{} between the two $T_1^{\text{spin}}$, which is roughly inline with our experimental values measured at \SI{6}{\kelvin} (MW-based control $T_1^{\text{spin}}=$\SI{2.5}{\milli\second}, all-optical control $T_1^{\text{spin}}=$\SI{1.65}{\micro\second}). We note that as a pre-factor, the temperature would not change the decay rate ratio between the two configurations. This ratio reiterates the fact that the Orbach process is the dominant factor for the $T_1^{\text{spin}}$ decay.

\section{MW control of the SnV spin}
\subsection{Randomized benchmarking}
The gates are chosen from the Clifford group and are \{$I$, $\pi_x$, $\pi_y$, $\pi_x/2$, $-\pi_x/2$, $\pi_y/2$, $-\pi_y/2$\}. We randomly choose ($N-1$)-gates and use the last gate to undo the sequence, followed by a z-basis measurement. The last gate is part of the Clifford group. 
We get the $\pi$-gate from Rabi measurements and adjust the time $t_\pi$ accordingly. The identity is implemented as wait-time for $t_\pi$, whereas $\pi/2$-gates have a duration of $t_{\pi/2}$. No buffer times are used which would make the qubit prone to dephasing errors, but the drive amplitude is reduced such that local heating effects is not a limiting factor. All randomized benchmarking measurements were taken at a Rabi frequency of $\Omega/(2\pi)=$ \SI{2.8}{\mega\hertz}. A total of 10 randomized sequences were applied each time to average out over different implementations. The fidelity $F$ is extracted by fitting the readout with $A*F^N+B$, from which we get the error per Clifford gate \cite{knill2008randomized}.

\subsection{Ramsey $T_{2*}$ at different qubit frequency}
\begin{figure}[H]
\centering
\includegraphics[width=0.8\textwidth]{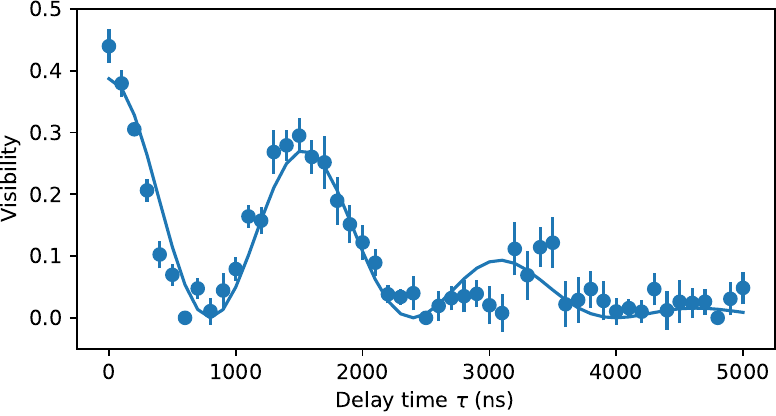}
\caption{Ramsey measurement of the inhomogeneous dephasing time $T_2^*$ at a qubit frequency of 3.694 GHz.}
\label{subfig:Ramsey117mT}
\end{figure}
We change the applied magnetic field magnitude from 81.5 to 117 mT and measure the Ramsey dephasing time $T_2^*$. We extract $T_2^*$ = \SI{2.63\pm0.14}{\micro\second} at the qubit frequency of 3.694 GHz (see Fig. \ref{subfig:Ramsey117mT}), indicating that $g$-factor fluctuations as reported in Ref.\cite{Sukachev2017} are not limiting the observed $T_2^*$.

\subsection{Ramsey measurements with phase-readout}
\begin{figure}[H]
\centering
\includegraphics[width=0.6\textwidth]{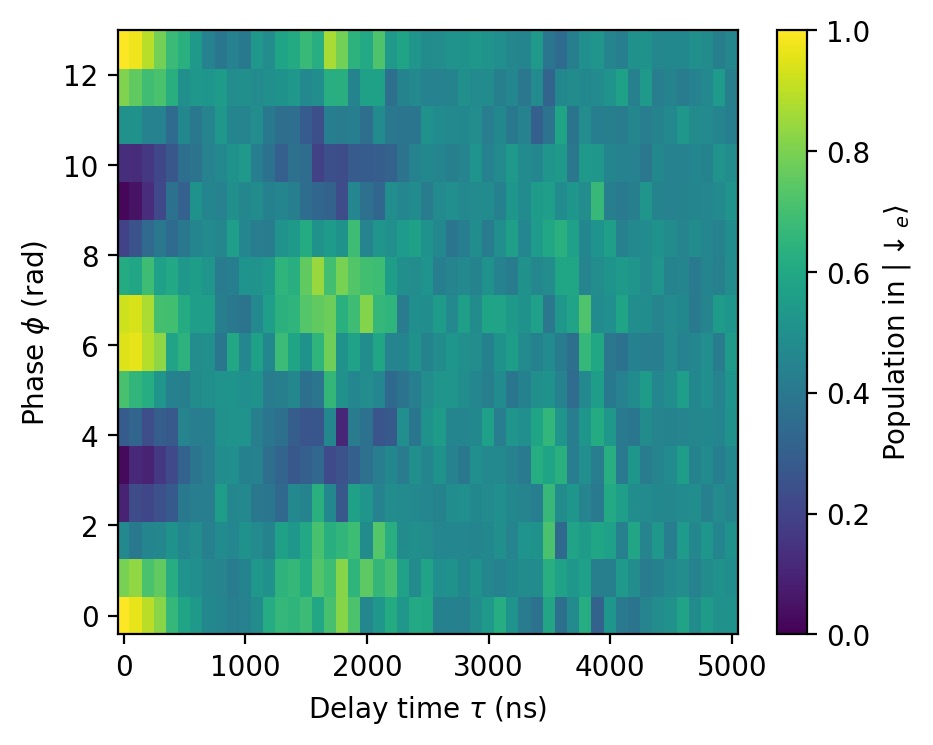}
\caption{Ramsey measurement of $T_2^*$. Both the delay-time $\tau$ and the phase of the second $\pi/2$-pulse are varied.}
\label{subfig:Ramsey}
\end{figure}

Fig. \ref{subfig:Ramsey} shows the measurement from which we extract $T_{2*}$ Ramsey and the modulation of the qubit frequency in Fig. 4 (a) in the main text. We fit for every time delay a modulation of the phase by $A*\cos{\phi}+B$, where $A$ is the visibility shown in the main text and $B$ is the mean value for all time delays averaged over all phases. We then extract the inhomogenous dephasing time $T_{2*}$ by fitting $A$ over delay time $\tau$ with an Gaussian envelope $\propto \exp{((-\tau/T_{2*})^2)}$. With this technique, we can distinguish with certainty a real modulation of the qubit frequency (loss in coherence and visibility)  versus a detuning error (no total loss of visibility, but no readout at certain delays and phases). The origin of the beating pattern needs further investigation. The MW-electronics were tested for any modulation. Likely candidates could be surface spins or substitutional nitrogen centres (P1-centres) with a large gyromagnetic ratio close to the one of a free electron ($g = 2$)  resulting in relatively large couplings even at large distances. 

\subsection{XY-sequences}
\begin{figure}[H]
\centering
\includegraphics[width=\textwidth]{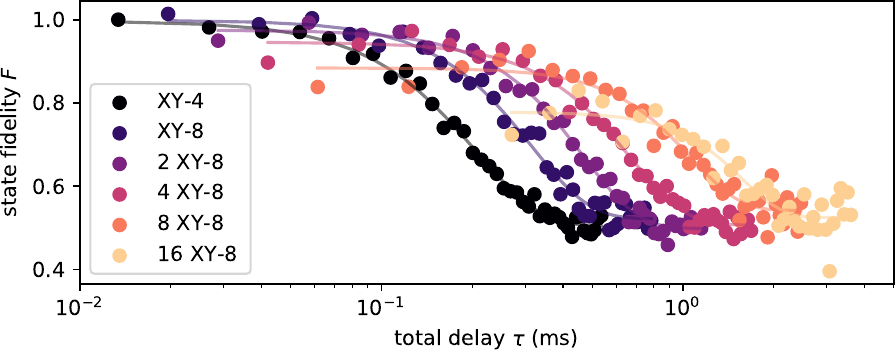}
\caption{State fidelity $F$ over total delay time $\tau$ for XY-pulse sequences.}
\label{subfig:XY}
\end{figure}

Fig. \ref{subfig:XY}  shows the coherence over total time delay $\tau$ in the XY-family of dynamical-decoupling sequences. We observe an almost identical scaling of coherence time over the number of $\pi$-pulses (see main text Fig. 4 c). Because of the much smaller sensitivity of XY- pulse sequences with regards to rotation- or offset errors compared to CPMG sequences, we conclude that our control does not limit the coherence \cite{suterDD}.

\section{Experimental details}
\subsection{Device fabrication}
The diamond membrane substrate was generated via \ch{He+} implantation and annealing. \SI{450}{\nano\meter}-thick diamond overgrowth layer was introduced in a diamond chemical vapor deposition (CVD) chamber, followed by a \ch{^{120}Sn} implantation with \SI{2e8}{\per\centi\meter\squared} dose and \SI{40}{\nano\meter} target depth. Individual membranes were patterned via lithography and electrochemically etched to undercut the graphitized layer. The target membrane was then transferred to a HSQ-coated fused silica substrate with a \SI{5}{\micro\meter}-deep etched trench to the generate suspended area. The substrate was annealed at \SI{600}{\celsius} for \SI{8}{\hour} in argon atmosphere. Membrane was thinned down to \SI{160}{\nano\meter} via ICP RIE etching using \ch{Ar/Cl2}, \ch{O2/Cl2}, and \ch{O2} recipes. The microwave coplanar waveguide was lithographically defined, followed by \ch{Ti} and \ch{Au} deposition with thicknesses of \SI{10}{\nano\meter} and \SI{200}{\nano\meter}, respectively. Excess resist was lift-off using \SI{80}{\celsius} NMP solutions.  

\subsection{Measurement setup}
All the measurement data in this work were taken in Cambridge, UK. 
The device was studied in a closed-cycle cryostat (attoDRY 2100) with a base temperature of \SI{1.7}{\kelvin} at the device and in which the temperature can be tuned with a resistive heater located under the sample mount. Superconducting coils around the sample space allow the application of a vertical magnetic field from 0 to \SI{9}{\tesla} and a horizontal magnetic field from 0 to \SI{1}{\tesla}. Unless explicitly stated otherwise, all measurements were conducted at $T=$ \SI{1.7}{\kelvin}. 
The optical part of the set-up consists of a confocal microscope mounted on top of the cryostat and a microscope objective with numerical aperture 0.82 inside the cryostat. The device is moved with respect to the objective utilizing piezoelectric stages (ANPx101/LT and ANPz101/LT) on top of which the device is mounted. Resonant excitation around \SI{619}{\nano\meter} is performed by a second harmonic generation stage (ADVR RSH-T0619-P13FSAL0) consisting of a frequency doubler crystal pumped by a \SI{1238}{\nano\meter} diode laser (Sacher Lasertechnik Lynx TEC 150). The frequency is continuously stabilized through feedback from a wavemeter (High Finesse WS/7). The charge environment of the SnV- is reset with microsecond pulses at \SI{532}{\nano\meter} (Roithner LaserTechnik CW532-100). PL measurements were done with a Teledyne Princeton Instruments PyLoN:400BR eXcelon CCD and SpectraPro HRS-750-SS Spectrograph. Optical pulses are generated with an acousto-optic modulator (Gooch and Housego 3080-15 in the \SI{532}{\nano\meter} path and AA Opto Electronics MT350-A0,2-VIS) controlled by a delay generator (Stanford Research Instruments DG645). For resonant excitation measurements, a long-pass filter at \SI{630}{\nano\meter} (Semrock BLP01-633R-25) is used to separate the fluorescence from the phonon-sideband from the laser light. The fluorescence is then sent to a single photon counting module (PerkinElmer SPCM-AQRH-16-FC), which generates TTL pulses sent to a time-to-digital converter (Swabian Timetagger20) triggered by an arbitrary waveform generator (Tektronix AWG70002A). Photon counts during ``initialize" and ``readout" pulses are histogrammed in the time-tagger to measure the spin-population. Sidebands driving both resonantly transitions as well as off-resonant all-optical control were generated by an amplitude electro-optic modulator (Jenoptik AM635), and the amplitude, phase, and frequency of the sidebands are controlled by a 25 Gs$/$sec arbitrary waveform generator (Tektronix AWG70002A). The EOM is locked to its interferometric minimum with a lock-in amplifier and PID (Red Pitaya, STEMlab 125-14) and using a freely available Lock-in+PID application \cite{Luda2019} with a feedback loop on the signal generated by a photodetector (Thorlabs PDA100A2). 

Microwave pulses are generated with the second channel of the arbitrary waveform generator and amplified with a low-noise amplifier (Minicircuits ZX60-83LN12+) and a high-power amplifier (Minicircuits ZVE-3W-83+). Microwave signals inside the cryostat are delivered via the in-built pico-coax cables, self-soldered cables and a customised PCB. The signal is transmitted through a second line and terminated outside of the cryostat with 50 Ohms. 

\begin{figure}[H]
\centering
\includegraphics[width=\textwidth]{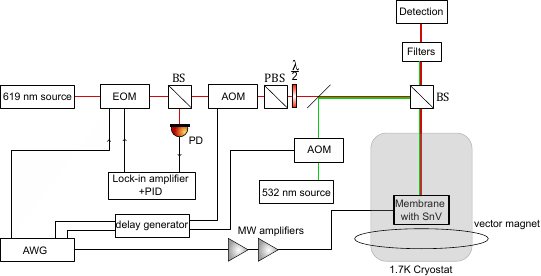}
\caption{Sketch of the experimental set-up described in detail in the text.}
\label{subfig:set-up}
\end{figure}

\bibliographystyle{Science}